\documentclass[reprint,superscriptaddress]{revtex4-2}
\usepackage{txfonts}
\usepackage{graphicx}
\usepackage{dcolumn}
\usepackage{bm}
\usepackage{hyperref}

\usepackage{amsmath}
\usepackage{amssymb}
\usepackage{moresize}
\usepackage{stmaryrd}
\usepackage{physics}
\usepackage{graphicx}  
\usepackage{subcaption}
\usepackage{xcolor}
\usepackage{bm}
\usepackage{float}  
\usepackage[left]{lineno}
\usepackage{soul}
\usepackage{xcolor}
\usepackage{array}
\usepackage{rotating}

\newcommand{\dimvec}{\text{dim}_{\text{vec}}}

\usepackage{verbatim}

\begin{document}

\preprint{APS/123-QED}

\title{
Experimental memory control in continuous variable \\ optical  quantum reservoir computing}

\author{Iris Paparelle}
\affiliation{Laboratoire Kastler Brossel, Sorbonne Universit\'e, ENS-Universit\'e PSL, CNRS, Coll\`ege de France, 4 place Jussieu, 75252 Paris, France}
\author{Johan Henaff}
\affiliation{Laboratoire Kastler Brossel, Sorbonne Universit\'e, ENS-Universit\'e PSL, CNRS, Coll\`ege de France, 4 place Jussieu, 75252 Paris, France}
\author{Jorge Garc{\'\i}a-Beni}
\affiliation{
 Instituto de F{\'\i}sica Interdisciplinar y Sistemas Complejos (IFISC), UIB–CSIC \\
 UIB Campus, Palma de Mallorca, E-07122, Spain}
\author{{\'E}milie Gillet}
\affiliation{Laboratoire Kastler Brossel, Sorbonne Universit\'e, ENS-Universit\'e PSL, CNRS, Coll\`ege de France, 4 place Jussieu, 75252 Paris, France}
\author{Daniel Montesinos}
\affiliation{
 Instituto de F{\'\i}sica Interdisciplinar y Sistemas Complejos (IFISC), UIB–CSIC \\
 UIB Campus, Palma de Mallorca, E-07122, Spain}
\author{Gian Luca Giorgi}
\affiliation{
 Instituto de F{\'\i}sica Interdisciplinar y Sistemas Complejos (IFISC), UIB–CSIC \\
 UIB Campus, Palma de Mallorca, E-07122, Spain}
\author{Miguel C. Soriano}
\affiliation{
 Instituto de F{\'\i}sica Interdisciplinar y Sistemas Complejos (IFISC), UIB–CSIC \\
 UIB Campus, Palma de Mallorca, E-07122, Spain}
\author{Roberta Zambrini}%
\affiliation{
 Instituto de F{\'\i}sica Interdisciplinar y Sistemas Complejos (IFISC), UIB–CSIC \\
 UIB Campus, Palma de Mallorca, E-07122, Spain}
\author{Valentina Parigi}
\email{valentina.parigi@lkb.upmc.fr}
\affiliation{Laboratoire Kastler Brossel, Sorbonne Universit\'e, ENS-Universit\'e PSL, CNRS, Coll\`ege de France, 4 place Jussieu, 75252 Paris, France}



\date{\today}

\begin{abstract}

Forecasting complex processes requires efficient learning from temporal data. Reservoir computing platforms enable such learning with minimal training cost. Quantum reservoir computing (QRC) extends this framework into the quantum domain, offering promising capabilities for online, quantum-enhanced machine learning tailored to temporal tasks. As in the classical case, photonics provides a natural platform for QRC. However, implementing native memory capabilities in practical photonic quantum systems remains a significant challenge.
Here, we demonstrate a photonic QRC platform based on deterministically generated multimode squeezed states, exploiting spectral and temporal multiplexing in a continuous-variable (CV) setting with controllable fading memory. Data is encoded via programmable pump phase shaping in an optical parametric process and retrieved through mode-selective homodyne detection. Real-time memory is implemented through feedback via electro-optic modulation, and expressivity is boosted via spatial multiplexing.
This architecture enables nonlinear temporal tasks, including parity check at different delays and chaotic signal forecasting. All results
are supported by a high-fidelity digital twin. Leveraging the entangled multimode structure enhances expressivity and memory capacity, establishing a scalable CV photonic platform for quantum-enhanced information processing.

\end{abstract}

\maketitle

\onecolumngrid



Reservoir computing is a machine learning neuro-inspired paradigm that exploits the dynamics of an artificial network with minimal learning requirements. Unlike conventional neural networks, it avoids the computational cost of tuning internal weights during training; instead, only the output layer is trained, typically via linear regression  \cite{lukovsevivcius2009reservoir,nakajima2021reservoir}.The untrained network, known as the reservoir, can be implemented as a physical process. 

Although the term reservoir computing is now broadly used for tasks involving an untrained network (i.e. a reservoir), its most distinctive feature is the ability to learn from temporal series (e.g., predicting chaotic time series)  by exploiting the memory properties of the system used as a reservoir. This temporal processing capability enables online learning and extends beyond static tasks such as classification traditionally addressed by architectures like Extreme Learning Machines
\cite{butcher2013reservoir,ortin2015unified}.

Quantum reservoir computing (QRC) extends the concept of physical reservoirs into the quantum regime and has recently been extensively explored across several platforms  \cite{Mujal21,ghosh2021quantum,labay2024neural,markovic} 
revealing both novel capabilities  \cite{Fujii17, Senanian2024,Nokkala2024, innocenti2023potential,GBeni2024Clust,kornjavca2024large,cimini2025}, and new challenges \cite{Mujal23mes,Sannia25}. Photonic platforms are particularly appealing 
as they inherit the capabilities of their classical counterparts \cite{Shastri21,McMahon2023}, offer many exploitable degrees of freedom, possess high integration potential, and, when information is encoded in quadrature continuous variables, can operate at room temperature using coherent detection for online monitoring \cite{GBeni2023}.  

Photonics systems have recently been tested in quantum machine learning tasks \cite{Yin25,Hoch25}.  In the context of QRC, following different proposals in either static \cite{sakurai2025quantum, nerenberg2025photon} and temporal tasks \cite{Nokkala2021,GBeni2023,Senanian2024},  some pioneering experimental results have been demonstrated on quantum optical platforms, these have primarily focused on classification tasks \cite{PhysRevLett.132.160802,zia2025,cimini2025} or classically assisted strategies \cite{PhysRevA.111.022609}. Only recently have efforts been directed at incorporating memory capabilities into general quantum platforms such as superconducting circuits \cite{Kobayashi24,monomi2025feedback,Hu2024}, and into a memristor-based photonic platform \cite{selimovic2025experimental}.

Here, we present an experimental realization of memory-tailored quantum reservoir computing using a reconfigurable, multimode continuous-variable photonic platform. Memory is engineered through feedback conditioned on the history of the quantum system, enabling online temporal learning.  Our approach exploits deterministically generated squeezed and reconfigurable entangled quantum states, multiplexed across both spectral and temporal modes \cite{Kouadou2023,Roman2024,Renault2023} along with reconfigurable mode-selective homodyne detection. The experimental realization
of temporal learning tasks via the continuous-variable builds on theoretical results that demonstrate the potential of such systems in QRC \cite{Nokkala2021,GBeni2023,GBeni2024,GBeni2024Clust}.
The experimental setup supports controllable memory demonstrating a performance boost in online learning and does not require post-processing with a classical neural network platform.

Data is encoded either via modulation of the global phase or of the complex spectral shape of the pump pulses driving the nonlinear optical process. Global phase modulation imprints information into the angle of the squeezing ellipses, tuning quantum correlations in the measurement basis. Spectral shaping directly modifies quantum correlations among output modes by altering the nonlinear interaction itself.

Moreover, we construct a digital twin of the experiment by integrating accurate models of the quantum and nonlinear optical processes, along with the continuous-variable dynamics of the reservoir. Once calibrated with a noise model, this simulation framework accurately reproduces experimental results and enables predictive exploration of the platform’s capabilities.

We experimentally perform temporal tasks enabled by memory, using global phase encoding. Memory is achieved via a feedback signal applied to the electro-optic phase modulator \cite{Henaff2024} that controls the pump phase; the feedback is derived from observables measured via mode-selective homodyne detection of the quantum reservoir’s multispectral components.
For short-memory tasks, we implement real-time feedback control, while longer-delay target functions are experimentally realized using a spatial multiplexing strategy, i.e., by running sequential experiments that act as spatially distributed copies of the quantum reservoir platform.

Finally we demonstrate that fully exploiting the entangled multimode nature of the system, yields better scaling of reservoir expressivity when compared to the use of multiple independent copies of the experiment (spatial multiplexing) and enhance the memory in temporal tasks.

This work demonstrates online temporal processing enhanced by controllable memory  using a continuous-variable (CV) entangled reservoir. Gaussian states and homodyne detection have been theoretically shown to increase information processing capacity compared to classical encodings \cite{Nokkala2021}, allowing saturation of a polynomial advantage as the reservoir size grows \cite{GBeni2023}. This regime is experimentally explored for the first time in this article, with potential implications for practical NISQ implementations. Furthermore, our architecture is compatible with measurement-based approaches using cluster states \cite{GBeni2024Clust} and with the integration of non-Gaussian resources \cite{Ra20,cimini2025}. 

While defining a quantum advantage in QRC, or more generally in quantum machine learning protocols, remains an open question \cite{Schuld22},  CV platform,  when augmented with non-Gaussian resources, are actively being pursued for fault-tolerant photonic quantum computing \cite{Bourassa2021}. Our demonstration of online temporal tasks boosted by memory opens the way to scalable quantum reservoir computing, as it requires only the deterministic generation of Gaussian states and deterministic room-temperature detection (e.g., homodyne). The platform provides a foundation for further exploration of quantum advantage using non-Gaussian resources \cite{Eaton2022,Winnel24,Anteneh24,larsen2025}.

\section{Results}

\subsection{The quantum resource}
\begin{figure}
    \centering
    \includegraphics[width=0.9\linewidth]{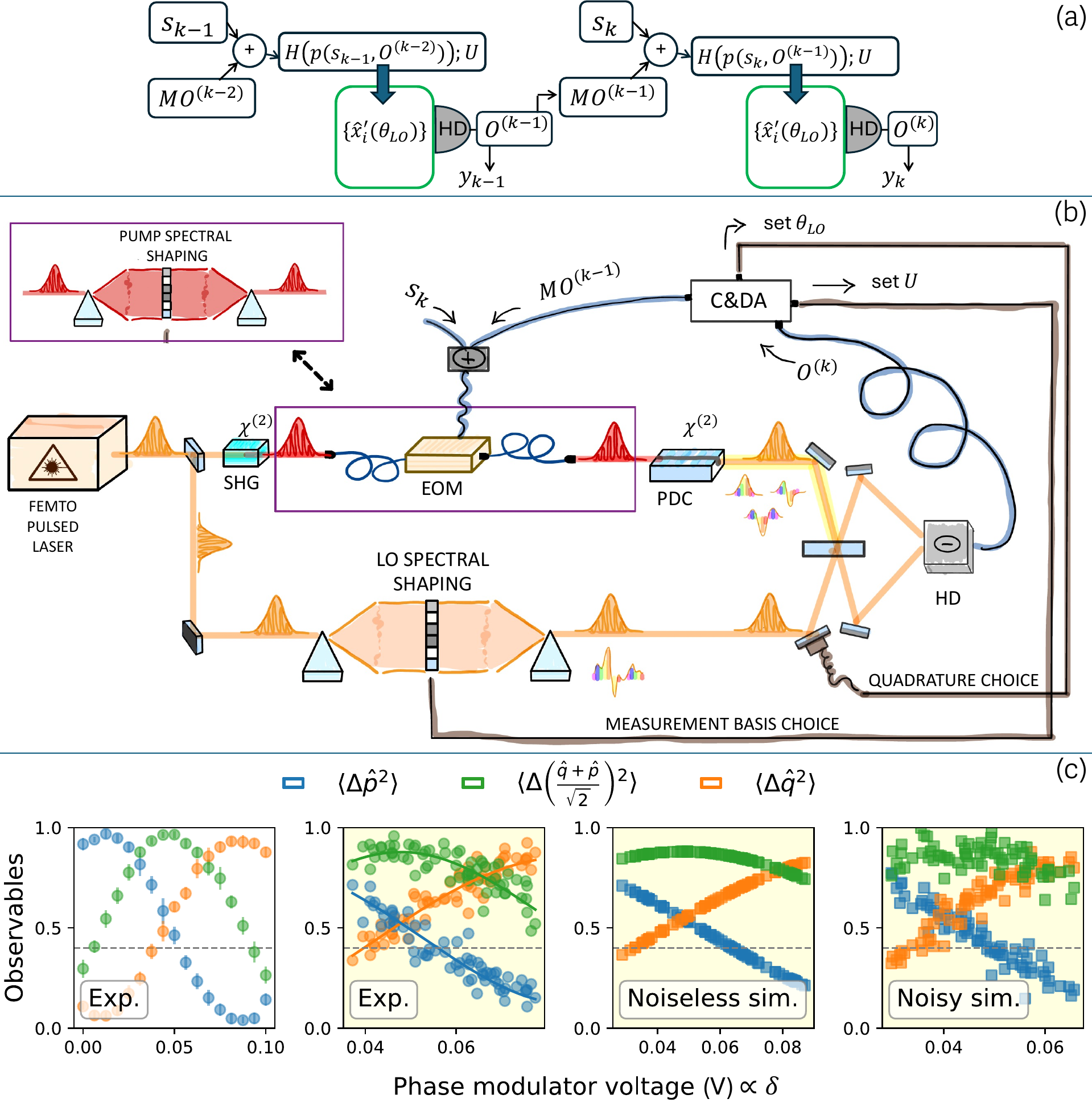}

    \caption{
(a) \textbf{Quantum reservoir protocol}. At each time step $k$, the input $s_k$ is added to a feedback signal derived from the previous observables $O^{(k-1)}$. This sets the pump of the parametric process, determining the operators $\hat{x}'_i$ and their quantum correlations. The task output $y_k$ is obtained from the observables. 
(b) \textbf{Experimental setup}. Information is encoded in the global phase of the pump, generated via second harmonic generation (SHG) and modulated by an electro-optic modulator (EOM). The more general encoding involve amplitude and phase shaping of the pump spectrum. The multimode entangled state from parametric down-conversion (PDC), and thus its observables, is tailored by this encoding. The control and data acquisition (C\&DA) module includes electronics to set the measurement basis ($U$) via local oscillator shaping in homodyne detection (HD), measure the observables $O_m$, and generate the feedback. Phase locking is not shown. 
(c) \textbf{Nonlinear dependence of observables} with the EOM voltage (proportional to the encoding phase $\delta$, which is a linear combination of $s_k$ and $O^{(k-1)}$}). Left: experimental data in low- and average-noise regimes. Right: simulated data using the Digital Twin model according to the experimental noise evaluation. The white panel spans a voltage range sufficient for observables to vary from 0 to 1 (phase difference slightly exceeding $\pi/2$). Yellow panels show observables recorded during training on a memory task.

    \label{fig:setup}
\end{figure}

\begin{figure}
    \centering
    \includegraphics[width=0.7\linewidth]{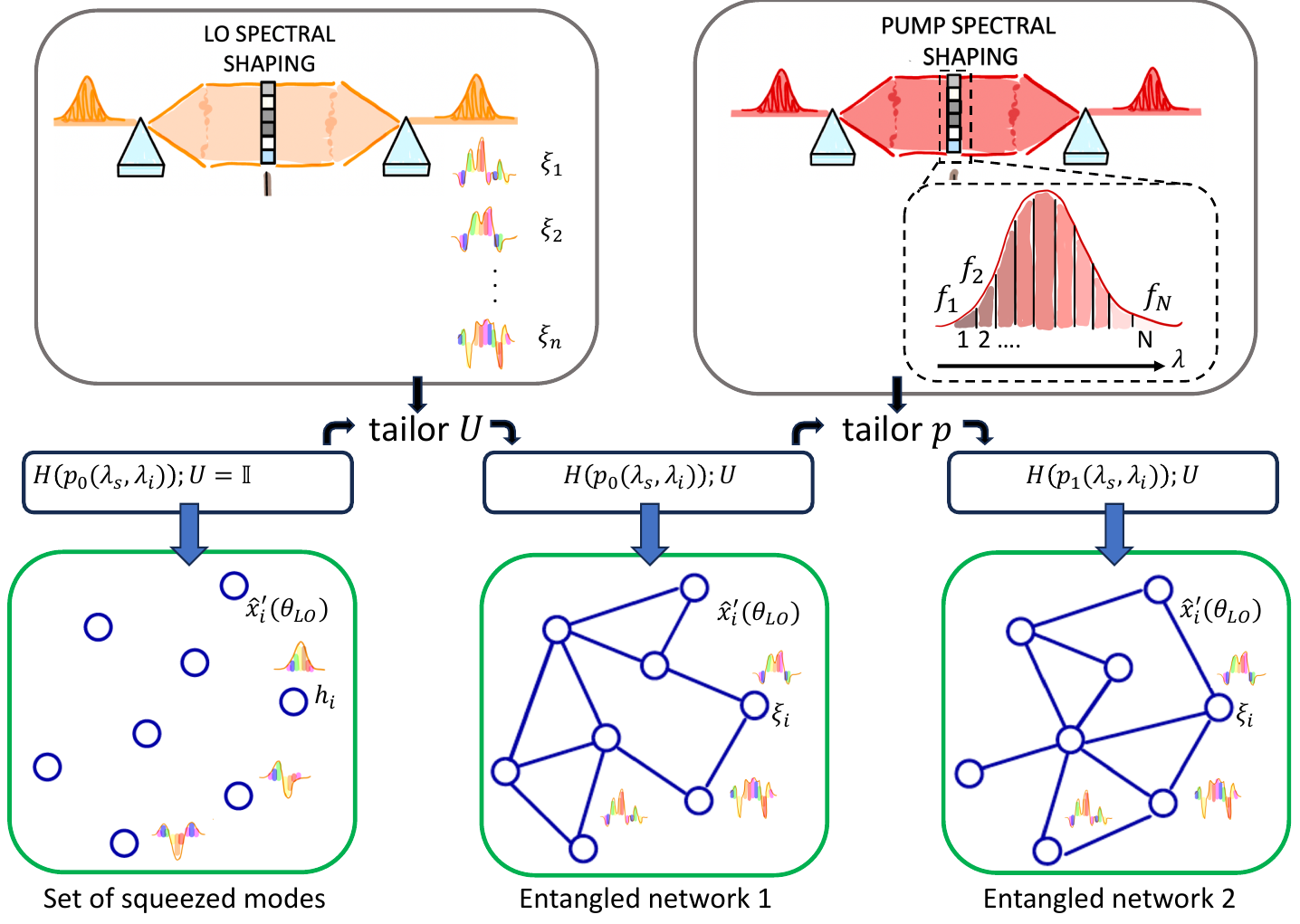}
    \caption{ \textbf{Controllability of the resource} by tailoring the local oscillator spectral shape and the pump spectral. The nodes of the network are different spectro-temporal modes of the field. The set of modes are defined by the basis choice $U$. For convention when $U = I $  the supermode basis $ \{h_i\}$ is chosen, and the modes are independently squeezed. In all the other mode bases $\{\xi_i\}$ the modes show entanglement correlations represented as links in a network. The general mode quadrature is $\hat{x}'_i(\theta_{LO})=\cos{(\theta_{LO})}\hat{q}_i'+\sin{(\theta_{LO})}\hat{p}_i'$ that corresponds to $\hat{q}_i'$ or  $ \hat{p}_i'$ if the local oscillator phase is set to be  $\theta_{LO}=0$ or $\theta_{LO}=\pi/2$. The correlations can be controlled by tailoring the pump of the process that modifies the Hamiltonian. 
     }
    \label{Fig:QR}
\end{figure}

The core quantum resource is a multimode optical network generated via parametric down conversion in a nonlinear waveguide pumped by a pulsed laser, as shown as in Fig.~\ref{fig:setup}. 
The interaction Hamiltonian describing the generation of signal and idler fields of wavelengths $\lambda_S,\lambda_I$ reads: \begin{equation}\label{eq:ham}
    \hat{H} \propto \int d\lambda_S d\lambda_I \, J(\lambda_S, \lambda_I) \, 
    \hat{a}^\dagger(\lambda_S) \hat{b}^\dagger(\lambda_I) + \text{h.c.}
\end{equation}
where  $J(\lambda_S, \lambda_I)$ is the so called 
joint spectral amplitude (JSA) which can be written as \cite{brecht2015photon,roman2021continuous,Arzani18}:
\begin{equation}\label{eq:JSA}
    J(\lambda_S, \lambda_I)=p(\lambda_S, \lambda_I)\cdot\phi(\lambda_S, \lambda_I)
\end{equation}

The term $p(\lambda_S, \lambda_I)$ is the complex pump spectral shape, while $\phi(\lambda_S, \lambda_I)$ is the phase-matching function, determined by the characteristics of the nonlinear medium.
The multimode nature of the generated quantum system can be retrieved by performing a Schmidt decomposition of the JSA $J(\lambda_S, \lambda_I)=\sum_k r_kh_k(\lambda_S)g_k(\lambda_I)$, where $h_k(\lambda_S)$ and $g_k(\lambda_I)$ are two sets of orthonormal modes, spectral profiles of signal and idler channels,
and  $r_k$ are called the Schmidt coefficients. 
In case of degenerate signal and idler fields, as in our experimental setup,   $  h_k=g_k; $ and the Hamiltonian in Eq.(\ref{eq:ham}) can be written as $\hat{H} \propto \sum_k r_k \, \hat{A}_k^{\dagger 2} + \text{h.c.}$, with $\hat{A}_k^\dagger $ broadband creation operators associated to the modes $ h_k$.  Acting on the vacuum, this Hamiltonian generates a tensor product of single-mode squeezed vacuum states, one in each mode $h_k(\lambda_S)$, that are also called supermodes.
The used setup can generate up to 40 Schmidt modes approximated by Hermite-Gauss polynomials as shown in Fig.~\ref{fig:setup}.

The produced quantum state is Gaussian and can be fully described by the covariance matrix $\boldsymbol{\sigma} $ of its quadratures, measured via homodyne detection.   The elements of the covariance matrix  are $\sigma_{ij}=1/2 \langle \Delta\hat{x}_i\Delta\hat{x}_j +\Delta\hat{x}_j\Delta\hat{x}_i \rangle$, being $\Delta\hat{x}_i = \hat{x}_i- \langle \hat{x}_i\rangle $ and $\hat{x}_i$ the quadrature $\hat{q}_i$ or $\hat{p}_i$ of the mode $i$ with commutation relation $ [\hat{q}_i,\hat{p}_i]=2 \imath $. 
The covariance matrix is obtained
by knowing the interaction Hamiltonian and, in the supermodes basis, takes the form of a diagonal matrix $\boldsymbol{\sigma}(H)=\text{diag}\{e^{2r_1}, e^{2r_2},...,e^{-2r_1}, e^{-2r_2},...\} $. Mode-selective homodyne detection allows us to probe the quantum state in arbitrary mode bases. If, instead of measuring the supermodes, we switch to another orthonormal mode basis $\{  \xi_j (\lambda_S) \}$, linked to the supermodes via the unitary transformation $\boldsymbol{U}$, the covariance matrix in the new basis becomes $\boldsymbol{\sigma}' = \boldsymbol{S_U}^{T} \, \boldsymbol{\sigma} \, \boldsymbol{S_U}$ with $\boldsymbol{S_U}$, a symplectic transformation driven by $\boldsymbol{U}$.  For more technical details, see Supplementary.

 The matrix $ \boldsymbol{\sigma}'$ is now non-diagonal and the off-diagonal terms are signatures of entanglement correlations between the measured modes.   Measuring on such basis allows to experimentally access  entanglement correlations for quantum protocols \cite{Cai2017,Renault2023}. The resource can then be pictured via an entangled network as shown in Fig.~\ref{Fig:QR}.
 
 The phase-matching function term of the JSA in Eq.(\ref{eq:JSA}) is fixed once and for all, as determined by the material and waveguide properties, and only the pump-envelope function can be modified in a reconfigurable way via mode-shaping techniques. This modifies the entanglement correlation between nodes as shown in Fig.~\ref{Fig:QR}.
 Therefore, we can explicitly write the dependence of the covariance matrix $\boldsymbol{\sigma}'$ on the pump $p$ and the chosen basis $U$ as $\boldsymbol{\sigma}'=\boldsymbol{S_U}^{T} \, \boldsymbol{\sigma(H(p)) } \, \boldsymbol{S_U}$.

In the following we exploit the control on the pump-envelope function to encode information in the quantum system in the configuration of a fixed measurement basis $U$. The expectation value of general observables $O_m=\langle\Delta\hat{x}_i\Delta\hat{x}_j\rangle$  that will be used in the reservoir task,  can be retrieved from the covariance matrix $\boldsymbol{\sigma}'(U,p)$.
Given the pump shape,  we have accurate numerical models of the process from the nonlinear interaction to the observables \cite{brecht2015photon,roman2021continuous,Arzani18}, which allows us to set a Digital Twin of the experimental setup as fully explained in Supplementary.




\subsection{Information encoding and memory control of the quantum reservoir}

When we tune the pump phase profile, the pump-envelope function can  be written as:
\begin{equation} p(\lambda_P, \boldsymbol{\delta}) = \sum_{i=1}^N  f_i(\lambda_P) e^{i\delta_i} \end{equation}
where the $f_i(\lambda_P)$, with $\lambda_P =(1/ \lambda_S +1/ \lambda_I)^{-1}$, are a set of $N$ pump modes, and the $\boldsymbol{\delta}=(\delta_i)_{i=1}^N\in \mathbb{R}^N$ are tunable phase parameters that allow precise shaping of the JSA to easily encode information. The pump input modes $f_i$ that we consider are $N$ frequency windows, delimiting a portion of our Gaussian pump profile $f(\lambda_P)$, that we can precisely control with a spatial light modulator for example. 
To perform the measurement, a set of $n$  modes  $\xi_j (\lambda_S) $ is selected by the basis choice $U$ via the local oscillator shaping in homodyne. The  tailoring of $N$ pump spectral components and the choice of the measurement basis composed by $n$ elements  are shown in the upper part of Fig.~\ref{Fig:QR}. 
More details about the input and output modes used are in Supplementary.

Our system is endowed with memory through a real-time feedback mechanism: the results of previous measurements are fed back into the system to influence future states. Specifically, if we consider a discrete-time sequence of inputs $\{s_k\}$, where $k$ denotes the timestep, the pump parameters are updated according to:
\begin{equation}
\label{eq:general_encoding}
        \delta_i^{(k)} = \alpha_i s_k + \beta_i + \sum_m M_{im}  O_m^{(k-1)}
\end{equation}
Here, $i\in1,...,N$, $\alpha_i$ and $\beta_i$ are real-valued coefficients, and $M_{im}$ is a feedback mask (a real matrix) that determines how previous measurement outcomes influence the current pump parameters. This determines the fading memory of the QRC. To summarize, $N$ determines the number of different input components that we encode in a single instance by manipulating several components of the pump spectrum: $N>1$ can be used to encode multidimensional inputs as well as to enrich the encoding of a scalar input, which will be then repeated across the pump spectrum with different parameters ($\alpha_i, \beta_i, M_{im}$). The quantities $O_m^{(k-1)} = \langle  \Delta\hat{x}_i \Delta\hat{x}_j  \rangle^{(k-1)}
$ are the values of the $m^\text{th}$ observable measured at timestep $k-1$, and   are derived from a subset of the elements of the covariance matrix ($i,j\in1,\dots,2n$). These are measured by accumulating data on the homodyne signal. We can then see that $n$ instead determines the size of the output layer of our reservoir, being the number of sections of the spectrum (modes) that we measure. 
So the reservoir computing is described  by the following:
\begin{equation}
    \begin{cases}
    O_m^{(k)} \propto \sigma'^{(k)}_{ij} \mid  \boldsymbol{\sigma}'^{(k)}= \boldsymbol{S_U}^{T} \, \boldsymbol{\sigma(H(p(s_k,\boldsymbol{O}^{(k-1)}))) } \, \boldsymbol{S_U}  \\

    y_k=\mathbf{w}^T \mathbf{O}^{(k)} + b

    \end{cases}
\end{equation}
 The reservoir is trained by adjusting the matrix $\mathbf{w}$ and parameter $b$  to minimize the error in the chosen learning task. The scheme of the quantum reservoir with feedback is shown in the top part of Fig.~\ref{fig:setup}. While we here focus on the case of feedback restricted to the last output at each cycle, the fading memory can be extended re-injecting a longer combination of past signals.

The case  $N>1$ allows to encode vector inputs $\vec{s}_k$, where $\dimvec(\vec{s}_k)\leq N$, by distributing the elements of $\vec{s}_k$ on the different phases $\delta_i$. 
A particular case of encoding corresponds to choosing $N=1$. In this case, we encode information on the global phase of the pump. This has been experimentally implemented by modulating the pump phase via an electro-optic modulator (EOM), as shown in Fig.~\ref{fig:setup}.

\subsection{Quantum reservoir computing tasks}

We evaluate the performance of our system across a range of benchmark tasks using two types of input encoding: the simple, experimentally implemented scheme with \( N = 1 \), and the more general encoding with \( N > 1 \). We begin with the XOR task, which assesses the system’s ability to capture nonlinearity and short-term memory while requiring minimal resources. We then consider the memory task to characterize the system’s fading memory property. Finally, we turn to more complex benchmarks, including the double-scroll and parity check tasks, which challenge the system’s scalability and its capacity to handle richer, higher-order nonlinear dynamics.

\subsubsection{
Experimental control of memory via global phase}
\begin{figure}
    \centering
    \includegraphics[width=\linewidth]{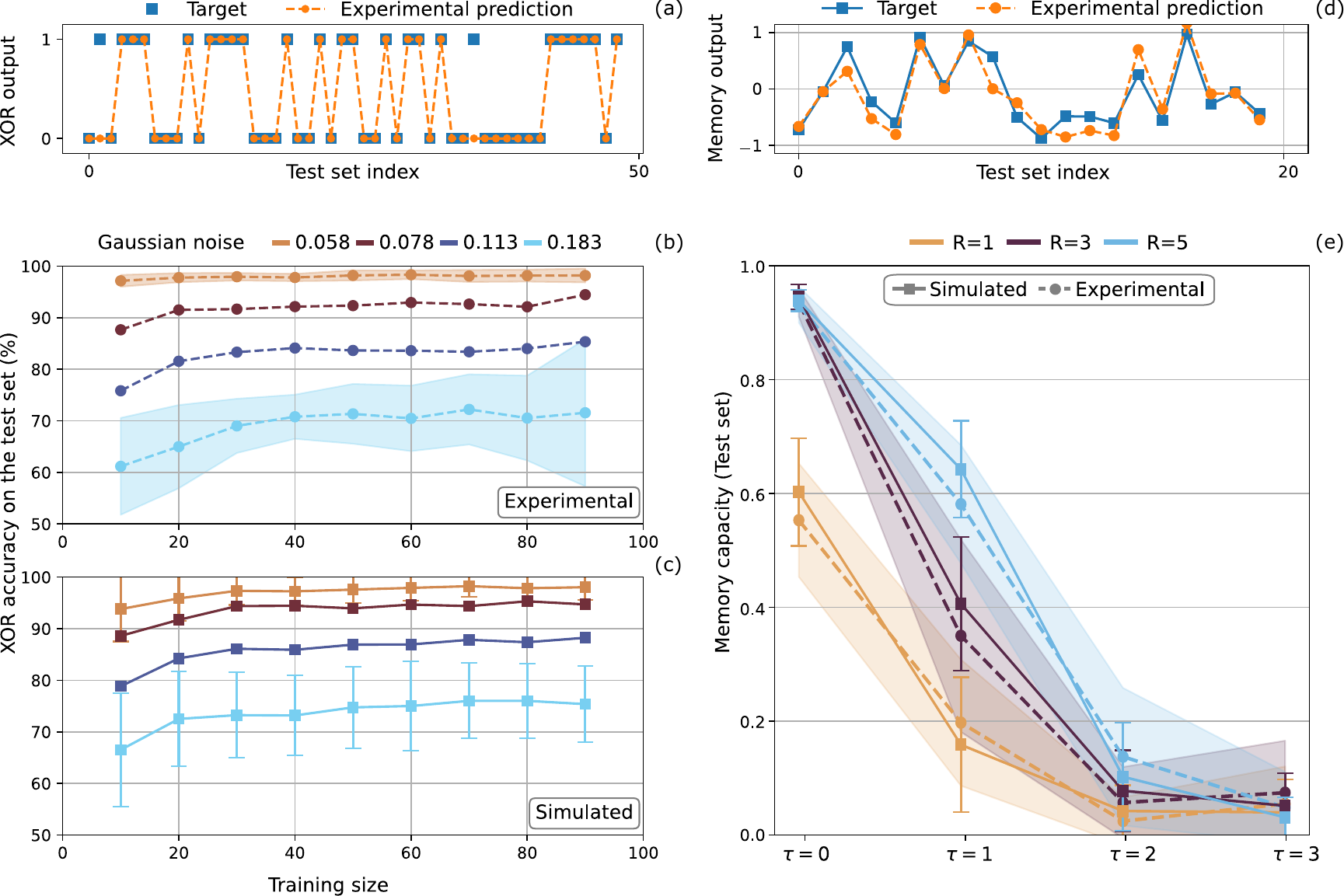}
\caption{\textbf{XOR task (experimental and simulated)}. (\textit{a}) Example of binary target (blue) and predicted (orange) outputs for the test set, with a test accuracy of $95.9\%$ (correct predictions over total). Training and test sizes are 99 and 49, respectively. (\textit{b,c}) Test accuracy versus training size: experimental (dots) and simulated (squares) results under varying noise levels (light brown, brown, blue, light blue), these noise levels are determined by calculating the standard deviation around the expected behavior of the observables (see the bottom of Fig.\ref{fig:setup}). Experimental error bars (shaded) are obtained by reshuffling data before the train/test split and calculating a standard deviation of the accuracy on these different realizations; simulated ones from repeated sampling. Larger training sets lead to smaller test sets, increasing variability due to single misclassifications. For clarity, error bars are shown only for the first and last noise levels. \textbf{Memory task (experimental and simulated)}. (\textit{d}) Continuous target (blue) and predicted (orange) output for a test set using $5$ reservoirs with delay $\tau=1$ (capacity $= 0.81$). (\textit{e}) Capacities vs. delay $\tau$: experimental (dots) and simulated (squares) results for different reservoir numbers (light brown, dark purple, light blue). Error bars are computed as for the XOR task. In both cases, simulations use a realistic experimental noise (for the memory task we had a Gaussian noise of 0.057)} model via the Digital Twin.

    \label{fig:exp}
\end{figure}

The phase encoding via the electro-optic modulator corresponds to the case of $N=1$, where the voltage applied changes linearly the pump phase $\delta$:
\begin{equation} p(\lambda_P, \delta) = f(\lambda_P) e^{i\delta} \end{equation}
where $f(\lambda_P)$ is the natural Gaussian profile of the pump.  As later shown in Subsec.\ref{sub:global}, with this type of encoding, the expressivity of the observables does not significantly increase with the number of measured modes $n$, therefore we use $n=1$ as well. This means that we directly measure by selecting the first supermode $h_1$ and we use as observables $\{ O_1, O_2, O_3 \}$ the set $  \{\langle\Delta \hat{q}^2\rangle$, $\langle\Delta \hat{p}^2\rangle,$ $\langle\Delta \hat{q}\Delta\hat{p}\rangle\sim \langle\Delta (\frac{\hat{q}+\hat{p}}  {\sqrt{2}})^2\rangle \}$, where we removed the mode index $1$ from the quadrature operators. These three observables are sufficient for implementing several tasks, as they express three different dependencies with respect to $\delta$ as shown in the lower part of Fig.~\ref{fig:setup}.
For linear regressions indeed, observables that are correlated (linearly dependent) are redundant (see \cite{appeltant2012reservoir, wijesinghe2019analysis}). For example, measuring $\langle\Delta (\frac{\hat{q}+\hat{p}}{\sqrt{2}})^2\rangle$ is equivalent to measuring $\langle\Delta \hat{q}\Delta\hat{p}\rangle$ since analytically, for a global phase encoding, $\langle\Delta \hat{q}\Delta\hat{p}\rangle(\delta)=  \langle\Delta (\frac{\hat{q}+\hat{p}}{\sqrt{2}})^2\rangle(\delta) +cst$ (as shown in Supplementary). 
In the bottom of Fig.~\ref{fig:setup} the first two panels show the experimental nonlinear dependence of the three observables  on $\delta$, encoded via the voltage of the phase modulator, in the case of a low-noise realization and a more noisy one, which is representative of the average noise in measurements.  Assuming a Gaussian noise model, the typical variance of observables is recovered via a Least Square method. The result is used in the Digital Twin to simulate the expected behavior as shown in the last two panels.

The unique non-zero term $M$ we use for the feedback mask is the one corresponding to $\langle\Delta \hat{q}^2\rangle^{(k-1)}$. At each timestep $k$, we set the global phase  depending on the input $s_k$ and the feedback as: 
\begin{equation}
    \delta^{(k)} = \alpha s_k + \beta + M \langle\Delta \hat{q}^2\rangle^{(k-1)}
\end{equation}
$M$, $\alpha$ and $\beta$ are randomly chosen within ranges that ensure both the input and the feedback contribute comparably to the phase. These ranges are selected to explore a nontrivial region of the observable space.

We begin with a binary task, the \textit{temporal XOR}, where the target output is $\hat{y}_k=s_k\bigoplus s_{k-1}$ and the input is a sequence of binary values $s_k\in \{0,1\}$. This task requires both nonlinearity as well as memory of the previous encoded input at each timestep. The reservoir is trained on a set of inputs using linear regression applied to the measured observables. The goal is to determine the optimal weights $\mathbf{w}, b$ that maximize accuracy (see Supplementary) by matching the predicted outputs $y_k$ (we put a threshold at $0.5$ to divide into $0$s and $1$s) with the known targets $\hat{y}_k$ in the training set.

The performance of the reservoir is then evaluated using the accuracy again, but on a test sequence of new data. In plot (a) of Fig.~\ref{fig:exp}, we present an example of the test target values and the corresponding experimental prediction. In plots (b) and (c) below, we show experimental and simulated accuracies as a function of training set size, for different levels of Gaussian noise. Experimental instabilities can introduce Gaussian noise in the measurement of the observables (see Fig.~\ref{fig:setup}), which can be mitigated by averaging over repeated measurements. These instabilities arise from mechanical and thermal fluctuations and are not intrinsic limitations of the protocol, but rather reflect the current performance of the experimental setup.
We see that the training saturates for set sizes around 100 and that noise affects the performance. Nevertheless, the reservoir could achieve experimentally test accuracies of $98\pm1\%$ for realistic levels of noise, with only $70$ training steps. The experimental noise is also well numerically simulated with a Gaussian model and the results are similar, as shown in plot (c) of Fig.~\ref{fig:exp}. 

A way to further test the memory of the system is through the \textit{linear memory} task, in which the reservoir is trained to reproduce past entries of the input series.  The target function at timestep $k$ is the input encoded $\tau$ steps in the past, $\hat{y}_k=s_{k-\tau}$. For the input, we generate a sequence of \textit{random} numbers uniformly distributed between 1 and -1, so $s_{k} \in [-1, 1]$. To do this task, the three previous observables are not sufficient and, as shown later, increasing the number $n$ of measurement modes does not help when encoding on a global phase. Therefore, we resort to spatial multiplexing, which is equivalent to use $R$ different reservoirs and therefore systems at the same time. This is a classical reservoir computing technique used to enhance the expressivity of the system, whereas the general encoding introduced in the next Section leverages spectral multiplexing, an intrinsic resource of the quantum system.

At each timestep, the same input $s_k$ is indeed fed to all $R$ reservoirs in parallel, each with their different parameters $\alpha_r$ and $\beta_r$ ($r\in1,\dots,R)$, this ensures that the observables of the $R$ reservoirs are not correlated. Concerning the feedback term, we collect the $R$ observables $\langle\Delta \hat{q}^2_{(r)}\rangle^{(k-1)}$ and we distribute them in the $R$ inputs with a mask $M$ of size $R\times R$. Each reservoir has then the following input at timestep $k$:
\begin{equation}
    \delta^{(k)}_{r}=\alpha_r s_k +\beta_r + \sum_{r'=1}^R M_{r,r'}\langle\Delta \hat{q}^2_{(r')}\rangle^{(k-1)} \quad r\in 1,...,R 
\end{equation}

The mask is a full matrix, meaning that the feedback term of each reservoir $r$ carries the observable values from all $R$ reservoirs. The output $y_k$ is then predicted using a linear regression to maximize the capacity (see Supplementary) on a training set. The performance of the reservoirs is then evaluated calculating the capacity again with the optimal weights found previously, but applied on a test set. In plot (d) of Fig.~\ref{fig:exp}, we present an example of target and experimentally predicted values. In plot (e), instead, we show the agreement between the experimental and simulated capacities on test sets when varying $\tau$ and the number $R$ of reservoirs. The simulations agree with the experimental results, confirming the necessity and advantage of multiplexing (better capacities for higher $R$) and the presence of a fading memory of the system (capacity gradually decreasing for higher delays). Experimentally, the parallelization of the process was implemented sequentially, sending the same input $R$ times to the physical setup but with different parameters for every repetition and collecting the observables to constitute the feedback term at the end of the cycle. 

\begin{figure}[h]
    \centering
    \includegraphics[width=\linewidth]{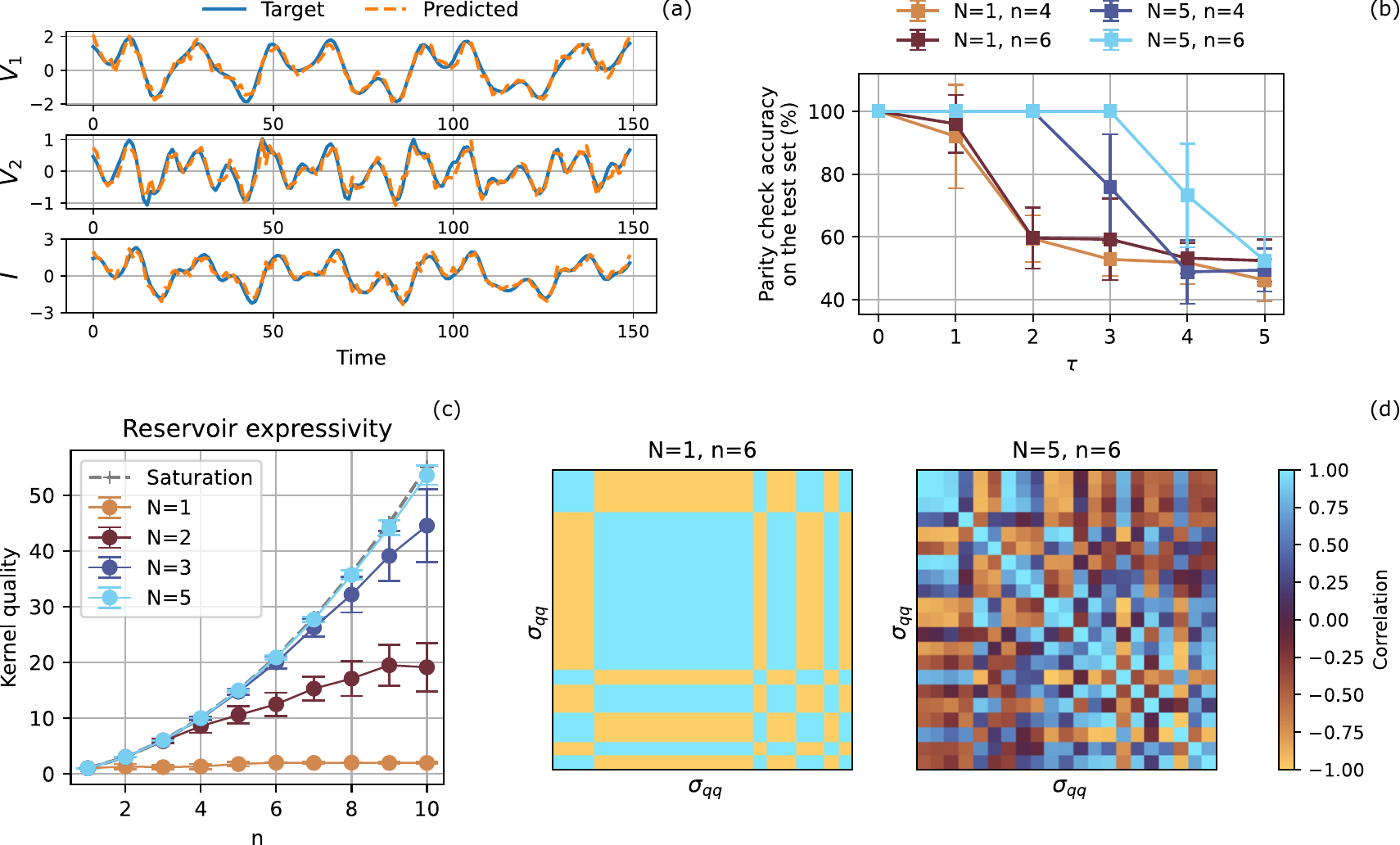}
\caption{\textbf{Double-scroll task (simulated)}. (\textit{a}) Target (blue) and predicted (orange) trajectories using a realistic experimental noise model (gaussian noise equal to $0.057$), with $15$ reservoirs and a training size of $350$.  Prediction accuracies for $V_1$, $V_2$, and $I$ are $93.1\%$, $85.3\%$, and $90.7\%$, respectively. \textbf{General encoding (simulated)}.(\textit{b}) Parity check accuracy for different delays $\tau$ and for different combinations of $N$ (the number of phase segments we imprint on the pump) and $n$ (the number of measured output modes). The error bars are obtained by repeating the simulation with random parameters $10$ times. (\textit{c}) Kernel quality (rank of the observables matrix) vs. number of measured modes $n$ (with $N$ input pump frexels), measuring reservoir expressivity. For $N=1$ (light brown), increasing $n$ does not improve expressivity, while for higher values of $N$ we get closer to the saturation of a polynomial  scaling with $n$ (quadratic in $n(n+1)/2$). Error bars reflect variations across random input sequences and parameters; fluctuations at $N=1$ mainly stem from numerical noise. (\textit{d}) Correlation matrices between observables for $N=1$ and $N=5$ at $n=6$ ($n$ $\hat{q}$-quadrature operators, yielding $n(n+1)/2=21$ observables). For $N=1$, we can see the high redundancy; for $N=5$, correlations and thus observables behaviors are richer and more diverse. When the rank of the observables matrix is $n(n+1)/2$, which is its dimension, it means every observable is meaningful and not redundant.}

    \label{fig:sim}
\end{figure}

Given the optimal agreement of the experiment with the numerical simulations, we use the Digital Twin to explore the performance of the protocol for a more complex task. Concretely, we perform forecasting of chaotic time-series such as  the \textit{double-scroll} electronic circuit \cite{Gauthier2021},  a three-dimensional dynamical system, and also (Supplementary) the \textit{Lorenz} system.
As we deal with $3$-dimensional input vectors, we again have to use spatial multiplexing at least to encode each of the three signals as inputs. Thus, we need a minimum of $3$ spatially multiplexed reservoirs to process the whole input. We can add more spatially multiplexed reservoirs to enhance the expressivity. 
For the training phase, we consider the target function to be the input one timestep in the future: $\mathbf{\hat{y}}_{k} = \mathbf{s}_{k+1}$. Once the system has learned to forecast the signal, we feed the reservoir predictions as new inputs and check if the system is able to reproduce the test set. Concerning the double-scroll, in plot (a) of Fig.~\ref{fig:sim}, we present an example of target and predicted values, obtained with simulations with realistic noise values. In Supplementary, we present three plots (corresponding to $V_1,V_2, I$) where the capacity is calculated while varying the training size and the number of reservoirs $R$. We prove that even with the experimental noise, we saturate the prediction capacity for relatively small values of $R$ (below 15) and of training size (below 400).

\subsubsection{
General encoding: scaling the expressivity and boosting the memory}\label{sub:global}

 When encoding on a global phase $\delta$, even if we increase the number of measured modes $n>1$, all the terms of the covariance matrix behave like either $\sin^2{\delta}$, $\cos^2{\delta}$ or $\cos{\delta}\sin{\delta}$. The expressivity of the reservoir is therefore limited for all $n$. This also happens when simulating the system with a detailed physical model that takes into account the characteristics of the parametric down conversion material  as shown in the bottom part of Fig.~\ref{fig:sim}, where we encode on a single pump band ($N=1$) and measure on $n$ output frequency bands (frexels). 
Indeed, to evaluate more quantitatively the expressivity of the system, we compute the rank of a matrix whose rows correspond to observables evaluated at different input states. 
This rank, called kernel quality in the literature \cite{appeltant2012reservoir,wijesinghe2019analysis, wringe2025reservoir}, would indicate the number of linearly independent observables, however since we scaled them beforehand, it also takes into account affine dependence. For linear regressions, an observable affine to another should not be useful. 
Since for a global phase, the observables behave like $\sin^2{\delta}=1-\cos^2{\delta}$, $\cos^2{\delta}$ or $\cos{\delta}\sin{\delta}$, analytically the rank should be equal to $2$ for all $n$, however numerical instabilities of the Schmidt decomposition yield small fluctuations. For example, we can see that there are two behaviors only ($O_A\sim\cos^2{\delta}$ and $O_B\sim-\cos^2{\delta}$) from the first subplot in (d) Fig.~\ref{fig:sim}, where we consider correlations in $\hat{q}$ quadratures alone.

To take advantage of the multimode nature of the system, it is necessary to diversify the phase profile of the pump ($N>1$). This means exploiting fully the phase encoding from Eq.(\ref{eq:general_encoding}), as shown in Fig.\ref{fig:sim}. In plot (b), with this general encoding ($N=5$), we achieve perfect capacities ($100\%$) for a parity check task for $\tau\in\{1,2,3\}$ and prove the boost in memory when  $n$, the number of measurement modes, is increased. Furthermore, as shown in plot (c), the expressivity increases with $n$ for $N>1$. Specifically, for $N>5$, we saturate a polynomial scaling with $n$ as reported in \cite{Nokkala2021}: the expressivity scales with $n(n+1)/2$, which is also the dimension of the observables matrix (with $n$ measurement modes and considering only $\hat{q}$ observables). This means that with a sufficiently large $N$ all the terms in our covariance matrix are independent. We highlight the increased diversity in behaviors of the observables in plot (d).  We also exploit the multidimensionality of the input (it can be a vector of size up to $N$), to implement again the double scroll task, without multiplexing. The performances of  the global phase encoding with multiplexing ($R=15$ reservoirs) and the general encoding with $N,n=9,9$ are similar, but the gain in scalability  is clear as  with the general encoding use only one system, instead of multiplexing in  $15$ reservoirs.

\section{Discussion}

Starting from a multimode squeezed resource, we develop an optical quantum reservoir computing platform that leverages spectral shaping capabilities in femtosecond optics for both encoding and decoding information.  Fading memory is enabled by linear feedback, even when limited to observables measured at the previous timestep. The continuous-variable (CV) setting is implemented via coherent homodyne detection, exploiting low-order quadrature moments as observables—an approach identified as offering polynomial advantage in QRC. 

By tailoring the pump phase in the parametric process using an electro-optic modulator, we experimentally show online temporal processing with memory, performing both binary tasks in real time and continuous tasks via successful spatial multiplexing. We demonstrate the nonlinear processing capabilities of the quantum reservoir with a temporal XOR task, assess its memory capacity using continuous input signals, and evaluate its forecasting performance with the double-scroll benchmark.
Results are reproduced by a Digital Twin, a detailed numerical model of the protocol that includes the properties of the nonlinear optical material, the optical processes, multimode and mode-selective detection schemes, and experimentally derived noise contributions. The Digital Twin enables exploration of the system’s capabilities with high fidelity.

The potential of the system’s multimode nature is further revealed by encoding information onto a richer phase profile of the pump. This allows consistent nonlinear processing, improved memory (as evidenced by a parity-check task), and accurate forecasting of chaotic time series, confirming the advantages of utilizing the full multimode structure. Compared to a purely spatial multiplexing strategy, this approach shows superior scaling, revealing the potential of the system’s entangled multimode structure for enhanced performance.

We also provide, in Supplementary Section J, a detailed comparison with classical machine learning approaches (Echo State Networks and Long Short-Term Memory networks) and with another QRC model. This analysis shows that our approach compares favorably with state-of-the-art time-series processing techniques, while relying on inexpensive training (linear regression), requiring only short training sequences, and exploiting polynomial scaling between the number of measured modes and the available observables. 

While polynomial scaling is achievable with Gaussian states, surpassing this performance necessitates the inclusion of non-Gaussian elements. Implementable candidates include spectral mode-selective photon subtraction, already demonstrated for spectrally multimode light fields \cite{Ra20}. 
Resources with stronger non-Gaussianity can be introduced via photon-number-resolving (PNR) detection in similar spectral mode-selective processes. In fact, PNR detectors have been used in integrated experimental setups to generate strongly non-Gaussian states such as Gottesman–Kitaev–Preskill (GKP) states \cite{larsen2025} . They have also been shown to enable nearly deterministic schemes for the generation of GKP  \cite{Anteneh24} and Schrödinger cat states \cite{Winnel24}. The experimental effort should go hand in hand with theoretical investigations of the advantages provided by non-Gaussianity in the continuous-variable (CV) setting, i.e., based on homodyne data readout. We can reasonably expect that even weak non-Gaussianity, introduced by heralding on a low number of detected photons, would enhance the effective nonlinearity of the platform.


\section{Methods}

\subsection{Experiment}
\label{sub:exp}

 The experimental setup is presented in Fig.~\ref{fig:setup}. We use a pulsed laser with a repetition rate of 100 MHz centered at 1560 nm, and a pulse duration of about 57 fs, which undergoes second harmonic generation in a periodically poled lithium niobate (ppLN) crystal to pump PDC in periodically poled potassium titanyl phosphate (ppKTP) waveguides of type 0. More details about the source can be found in  \cite{Roman2024}. In the setup of this work, an electro-optic modulator (iXblue NIR-800-LN-0.1) is used to set the phase of the pump beam for the global phase encoding.
 Phase lock is provided in order to encode information and measure the desired quadratures at the corresponding phases. The general multi-spectral phase encoding can be set via a pulse-shaping technique based on a spatial light modulator, as it is implemented in the homodyne setup to select the spectral basis of the measurement. 
The average value of squeezing measured in this setup is  $-0.45$ dB, a reduced value compared to the one in \cite{Roman2024} due to the dispersion introduced in the pump pulse by the fiber-based phase modulator. Such dispersion can be significantly reduced in future experiments by shortening the input and output fiber lengths. The antisqueezing is around $0.7$ dB because of losses.

\subsection{Simulations - Digital Twin}

Our experiment is fully characterized and accurately reproducible via simulation. This allows for faithful reproduction of experimental results and reliable predictions for new configurations, including the impact of additions such as a spatial light modulator on the pump. Although simulations are much slower than the femtosecond-scale nonlinear processes in the waveguides, they constitute a useful Digital Twin of the experiment, enabling interpretation and hypothesis testing. 

We begin by modeling the nonlinear parametric down-conversion (PDC) process in the waveguides. This involves computing the joint spectral amplitude (JSA), determined by the phasematching function - derived from the waveguide geometry and material (via the Sellmeier equations) -  and by the complex pump beam profile, which in our protocols encodes the input information. From the JSA, we extract the Schmidt modes and squeezing coefficients via singular value decomposition (SVD) - the most computationally intensive step. We model the measurement stage by computing overlaps between Schmidt and measurement modes, which allows us to finally reconstruct expected observables for arbitrary pump phase profiles. The core of the simulations stems from a detailed physical model shown in \cite{roman2021continuous} and implemented in a Python code.

\begin{acknowledgments}
This   work   was   supported   by   the   European   Research Council under the Consolidator Grant COQCOoN (Grant No.  820079) and by Agence Nationale de la Recherche (OQuLus, ANR-22-PETQ-0013).
We acknowledge the Spanish State Research Agency, through the Mar\'ia de Maeztu project CEX2021-001164-M, through the COQUSY project PID2022-140506NB-C21 and -C22, through the INFOLANET project PID2022-139409NB-I00, and through the QuantCom project CNS2024-154720, all funded by MCIU/AEI/10.13039/501100011033; the project is funded under the Quantera II programme that has received funding from the EU's H2020 research and innovation programme under the GA No 101017733, and from the Spanish State Research Agency,  PCI2024-153410 funded by MCIU/ AEI/10.13039/50110001103; MINECO through the QUANTUM SPAIN project, and EU through the RTRP - NextGenerationEU within the framework of the Digital Spain 2025 Agenda; CSIC’s Quantum Technologies Platform (QTEP). J.G-B. is funded by the Conselleria d’Educació, Universitat i Recerca of the Government of the Balearic Islands with grant code FPI/036/2020 and D.M. by the ”la Caixa” Foundation (ID 100010434), code LCF/BQ/DR24/12080033.

\end{acknowledgments}



\section{Supplementary}

\subsection{Simulations - Digital Twin}

The simulation is based on the effective interaction Hamiltonian for the Parametric Down Conversion (PDC) process, which reads, as introduced in the main text:
\begin{equation}
    \hat{H} \propto \int d\lambda_S \, d\lambda_I \, J(\lambda_S, \lambda_I) \, \hat{a}^\dagger(\lambda_S) \hat{b}^\dagger(\lambda_I) + \text{h.c.}
\end{equation}
where \( J(\lambda_S, \lambda_I) \) is the Joint Spectral Amplitude (JSA), given by the product of the complex pump envelope \( p(\lambda_S, \lambda_I) \) and the phase-matching function \( \phi(\lambda_S, \lambda_I) \). The complex pump function $ p(\lambda_S, \lambda_I)$  can actually be written as $p(\lambda_p)$ because of energy conservation condition $\lambda_p= (1/\lambda_S + 1/\lambda_I )^{-1}$. It conveys the information in our protocols and is described in detail in Subsec.~\ref{subsec:n}. For the pump, we employ a Gaussian amplitude with a phase profile segmented into $N$ sections, each assigned a distinct phase value. If $N=1$ we are applying a global phase $\delta$, such as we do experimentally with our electro-optic modulator (EOM). Instead, for $N>1$ we need a spatial light modulator, to encode a vector phase $\boldsymbol{\delta}=(\delta_i)_{i=1}^N$ where each element is the phase of the corresponding frequency band of the pump. 
The phase-matching function instead captures the effect of dispersion and waveguide geometry and is computed as:
\begin{equation}
    \phi(\lambda_S, \lambda_I) = \text{sinc}\left( \frac{L \, \Delta k(\lambda_S, \lambda_I)}{2} \right),
\end{equation}
with \( L \) the length of the nonlinear poled waveguide (periodically poled potassium titanyl phosphate - ppKTP) and \( \Delta k \) the phase mismatch given by:
\begin{equation}
    \Delta k(\lambda_S, \lambda_I) = k_P\left( \lambda_P \right) - k_S(\lambda_S) - k_I(\lambda_I) -\frac{2\pi}{\Lambda},
\end{equation}
where \( \lambda_P^{-1} = \lambda_S^{-1} + \lambda_I^{-1} \) ensures energy conservation and $\Lambda$ is the poling period of the waveguide, calculated in order to minimize the mismatch around the central wavelengths of the experiment. Each wavenumber is given by \( k_X(\lambda) = \frac{2\pi n_X(\lambda)}{\lambda} \), where \( X \in \{P, S, I\} \), and \( n_X(\lambda) \) is the effective refractive index, which depends on the material, polarization, and waveguide dimensions.

In our simulations, we use the Sellmeier equations for KTP, our biaxial crystal, to model the wavelength-dependent refractive indices along the principal axes \( x \), \( y \), and \( z \). These equations take the empirical form:
\begin{align}
    n_x^2(\lambda) &= X_1 + \frac{X_2}{\lambda^2 - X_3} + \frac{X_4}{\lambda^2 - X_5}, \\
    n_y^2(\lambda) &= Y_1 + \frac{Y_2}{\lambda^2 - Y_3} + \frac{Y_4}{\lambda^2 - Y_5}, \\
    n_z^2(\lambda) &= Z_1 + \frac{Z_2}{\lambda^2 - Z_3} + \frac{Z_4}{\lambda^2 - Z_5},
\end{align}
where the coefficients \( X_i \), \( Y_i \), and \( Z_i \) are experimentally determined constants taken from Ref.~\cite{kato2002sellmeier}. Corrections are also included to account for the effect of waveguide confinement, which modifies the effective refractive index depending on the mode dimensions. The polarization of the fields (horizontal or vertical) is associated with the appropriate axis depending on the phase-matching type (Type-0, I, or II), which determines whether the pump, signal, and idler share the same or different polarization states. In our experiment, we use Type-0 waveguides. 

To simulate the quantum output, we perform a singular value decomposition (SVD) of the JSA:
\begin{equation}
    J(\lambda_S, \lambda_I) = \sum_k r_k \, h_k(\lambda_S) \, g_k(\lambda_I).
\end{equation}
This yields two orthonormal sets of spectral modes \( \{h_k\} \) and \( \{g_k\} \), associated with the signal and idler fields, respectively, and a corresponding set of Schmidt coefficients \( \{r_k\} \). These modes define a new broadband mode basis through the operators:
\begin{equation}
    \hat{A}_k^\dagger = \int d\lambda_S \, h_k(\lambda_S) \, \hat{a}^\dagger(\lambda_S), \quad
    \hat{B}_k^\dagger = \int d\lambda_I \, g_k(\lambda_I) \, \hat{b}^\dagger(\lambda_I).
\end{equation}

In the degenerate case relevant to our experiment, where signal and idler are indistinguishable, we have \( h_k = g_k \) and \( \hat{A}_k^\dagger = \hat{B}_k^\dagger \), and the Hamiltonian simplifies to:
\begin{equation}
    \hat{H} \propto \sum_k r_k \, \hat{A}_k^{\dagger 2} + \text{h.c.}
\end{equation}
confirming the multimode squeezed output state. 

To model the effect of measurement in a different basis (see Subsec. \ref{subsec:n} for more details), we construct a mode transformation matrix \( U \) that maps the natural Schmidt basis of the photon pair source to user-defined measurement bins. Formally the new mode basis basis $\xi_j (\lambda_S)$ with creation operators $\hat{C}_j^\dagger$  is given by  $\{\hat{C}_j^\dagger=\sum_{j}U_{jk}\hat{A}_k^\dagger ; \; \xi_j (\lambda_S)=  \sum_{j}U_{jk} h_k(\lambda_S) \}$. Each element of \( U \) is computed by integrating a Schmidt mode over the spectral range of a given frequency band (frexel) using the midpoint rule. The columns of \( U \) are then normalized.
To simulate quadrature measurements, we construct the associated symplectic matrix $\boldsymbol{S_U}$, which operates on the covariance matrix in the grouped ordering \([q_1, \ldots, q_{n}, p_1, \ldots, p_{n}]\). The matrix $\boldsymbol{S_U}$ is built from the real and imaginary parts of \( U \) as:
\begin{equation}
 \boldsymbol{S_U} = \begin{pmatrix}
\mathrm{Re}(U) & -\mathrm{Im}(U) \\
\mathrm{Im}(U) & \mathrm{Re}(U)
\end{pmatrix}.   
\end{equation}
The covariance matrix in the measurement basis is then obtained via a congruence transformation:
\begin{equation}
    \sigma_\text{frexel} = \boldsymbol{S_U}^\mathrm{T} \, \text{diag}\{e^{2r_1}, e^{2r_2},...,e^{2r_n}, e^{-2r_1}, e^{-2r_2},...,e^{-2r_n}\}  \, \boldsymbol{S_U}= \boldsymbol{\sigma}'(U,p),
\end{equation}

 Our observables correspond to a subset of the elements in $\sigma_{\text{frexel}}$ since while they depend nonlinearly on the pump phase profile, they can be affine functions of each other, leading to redundancy.

For the case \( N = 1 \), where the information is encoded in the global phase of the pump, analytical expressions for the observables are available (see Subsec.~\ref{sub:analytical}). In this regime, using the full simulation pipeline would be excessive, so we instead implemented a simplified model based directly on the analytical formulas, with the addition of Gaussian noise, with equivalent results. The noise parameters were estimated by fitting to the experimental data using the least-squares method. We also adjusted the squeezing parameters (\( r_k \)) and the phase dependence to match the experimental values and results. These simulations were also useful to establish the range of parameters $\alpha, \beta, M$ used in the experiment to encode information on the phase.

\subsection{$N$ input modes and $n$ output modes}
\label{subsec:n}

Let us consider a Gaussian pump profile \( f(\lambda_P) \) of width \( \sigma \) and central wavelength \( \lambda_P^c \). Let \( N \) be the dimension of our input space, that is to say the number of pump frequency modes.

The frequency modes \( f_i(\lambda_P) \) ($i=1,\dots,N$) considered in the simulations - and implementable using a spatial light modulator - are defined as:
\begin{equation}
f_i(\lambda_P) =
\begin{cases}
 f(\lambda_P) & \text{if } \lambda_P \in \left[\lambda_P^c - 3\sigma + 6\sigma \frac{i}{N}, \; \lambda_P^c - 3\sigma + 6\sigma \frac{i+1}{N}\right] \\
0 & \text{otherwise}
\end{cases}
\end{equation}

In other words, the total \( 6\sigma \)-wide Gaussian pump is divided into \( N \) contiguous, non-overlapping frequency segments (frexels).

Concerning the \( n \) measurement modes, for the simulations with the general encoding ($N>1$) we use as measurement modes frequency bands (frexels again). If the Schmidt basis is centered around \( \lambda^c = 2\lambda_P^c \), we consider an interval \( [-S, S] \) centered on \( \lambda^c \) such that it includes the main features of the first \( n \) Schmidt modes. 

The frexels are defined as:

\begin{equation}
\xi_j(\lambda) =
\begin{cases}
1 & \text{if } \lambda \in \left[\lambda^c - S + \frac{2Sj}{n},\; \lambda^c - S + \frac{2S(j+1)}{n}\right] \\
0 & \text{otherwise}
\end{cases}
\end{equation}

That is, each measurement mode corresponds to a rectangular frequency band and together the \( n \) modes form a uniform partition of the interval \( [\lambda^c - S,\; \lambda^c + S] \), covering the spectral region where the Schmidt modes are significantly supported. This can be easily implemented again with the spatial light modulator on the local oscillator. Here, the amplitude is set to $1$ because in the simulations everything is normalized after the calculation of the JSA.

The effect of this choice of measurement modes is calculated in the matrix \( U \), where the $h_i(\lambda)$ are the Schmidt modes:
\begin{equation}
U_{ij} = \int h_i(\lambda) \, \xi_j(\lambda) \, d\lambda
\end{equation}
When each measurement mode \( \xi_j(\lambda) \) is a frexel this simplifies to:
\begin{equation}
U_{ij} = \int_{\lambda^c - S + \frac{2Sj}{n}}^{\lambda^c - S + \frac{2S(j+1)}{n}} h_i(\lambda) \, d\lambda 
\end{equation}

That is, \( U_{ij} \) quantifies the overlap between the \( i \)-th Schmidt mode and the \( j \)-th measurement frequency band. The matrix \( U \) therefore describes how the original modal structure is projected onto the coarse-grained measurement basis and enables the computation of the corresponding covariance matrix in simulations. The rows of \( U \) are normalized to preserve unitarity as closely as possible. However, since the Schmidt modes form a complete (infinite-dimensional) basis and the measurement modes form a finite set—even if they are orthogonal and span the same subspace—the matrix \( U \) does not represent an exact change of basis. Rather, it provides a truncated, approximate mapping suitable for finite-dimensional analysis.


\subsection{Regressions and metrics}

To train a reservoir computer we use regressions to combine the observables in a way that mimics a target function. Indeed, we compare a predicted output $y$ with a target output $\hat{y}$, of length $M$.

For binary tasks, where each prediction is either 0 or 1, we use a linear regression initially: 
\begin{equation}
    y_k = \mathbf{w}^T \mathbf{O}^{(k)} + b
\end{equation}
However, before evaluating the performance, we apply a threshold at 0.5 to obtain binary predictions:
\begin{equation}
    {y'}_k = 
    \begin{cases}
        1 & \text{if } y_k > 0.5, \\
        0 & \text{otherwise}.
    \end{cases}
\end{equation}
We evaluate the performance of the system with the accuracy, defined as the fraction of correct predictions:
\begin{equation}
    \text{Accuracy} = \frac{1}{M} \sum_{i=1}^M \delta\left( y'_i, \hat{y}_i \right)
\end{equation}
where $\delta(a, b)$ is the Kronecker delta function, equal to 1 if $a = b$ and 0 otherwise.

For continuous tasks, we use a linear regression directly: 
\begin{equation}
    y_k=\mathbf{w}^T \mathbf{O}^{(k)} + b
\end{equation}
To evaluate the performance, we use here the capacity, which is defined as the squared Pearson correlation coefficient between the predicted output $y$ and the target output $\hat{y}$ (as in \cite{martinez2021dynamical}):
\begin{equation}
    \text{Capacity} = \left( \text{corr}(y, \hat{y}) \right)^2 = \left( \frac{\text{Cov}(y, \hat{y})}{\sigma_{y} \sigma_{\hat{y}}} \right)^2
\end{equation}

Additionally, we report the normalized mean squared error (NMSE), which is widely used in the literature for benchmarking reservoir computing models. The NMSE measures the average squared deviation between the predicted and target signals, normalized by the mean squared value of the target:
\begin{equation}
    \text{NMSE} = \frac{\langle (y - \hat{y})^2 \rangle}{\langle y^2 \rangle}  ,
\end{equation}

Concerning the expressivity, let $\sigma'_k$ be the covariance matrix obtained at timestep $k$ from the encoded input $s_k$ and feedback in the measurement basis.
We construct a matrix $\mathcal{O} \in \mathbb{R}^{d \times d}$, where each row contains the elements of the covariance (there are at most $d=n(2n+1)$ unique elements) for $d$ steps after some washout inputs. We scale the observables and we define:
\[
 \mathrm{rank}(\mathcal{O})= \#\{\lambda_i > \varepsilon\ |\ \lambda_i \text{ is a singular value of } \mathcal{O}\} = \text{kernel quality} 
\]
where the rank is computed with a numerical tolerance $\varepsilon = 10^{-4}$. This rank, often called kernel quality, indicates how well the reservoir represents different input streams
and it can be used as a measure for the complexity and diversity of nonlinear
operations performed, that is to say the expressivity of the reservoir \cite{appeltant2012reservoir, wijesinghe2019analysis, wringe2025reservoir}. The rank defines the number of observables that are not linearly dependent (after scaling).

\subsection{Analytical expressions of the global phase encoding} \label{sub:analytical}
Let us see what happens when measuring more than one mode, that is to say, $n>1$, for the global phase encoding. The set of the $n$ measurement modes is related to the Schmidt modes by $U$. To simplify the notations, we restrict ourselves to the case where $U$ is real.

The covariance matrix of the system becomes:

\begin{equation}
\sigma(\delta, U)=
\begin{bmatrix} 
    \sigma_{0,0}(\delta, U) & \sigma_{0,1}(\delta, U) & \dots \\
    \vdots & \ddots & \\
    \sigma_{n-1, 0}(\delta, U) &        & \sigma_{n-1, n-1} (\delta, U)
\end{bmatrix}
\end{equation}

Where $\sigma_{l,j}(\delta, U)$ is a 2x2 matrix:

\begin{equation}
\sigma_{l,j}(\delta, U)=\sum_{a=1}^{n} U_{j,a} U_{l,a} \begin{bmatrix} 
    e^{2r_a}+(e^{-2r_a}-e^{2r_a})\sin^2{\delta}& (e^{2r_a}-e^{-2r_a})\cos{\delta}\sin{\delta}\\
    (e^{2r_a}-e^{-2r_a})\cos{\delta}\sin{\delta} & e^{2r_a}+(e^{-2r_a}-e^{2r_a})\cos^2{\delta}
\end{bmatrix}    
\label{eq:analytic}
\end{equation}

We see that all the terms of the covariance matrix are affine functions of either \( \sin^2{\delta} \), \( \cos^2{\delta} \), or \( \cos{\delta}\sin{\delta} \), reflecting the phase dependence introduced by the rotation in phase space.

Let us restrict ourselves to the case $n=1$ for simplicity, the rotated quadrature variance along the direction \( (\hat{q} + \hat{p})/\sqrt{2} \) is given by
\begin{equation}
    \langle\Delta (\frac{\hat{q}+\hat{p}}  {\sqrt{2}})^2\rangle 
= \frac{1}{2}\left( \langle\Delta \hat{q}^2\rangle + \langle\Delta \hat{p}^2\rangle + 2\langle\Delta \hat{q} \Delta\hat{p}\rangle \right)=\frac{1}{2}(2e^{-2r}+2(e^{2r}-e^{-2r})\cos\delta\sin\delta)=\langle\Delta \hat{q} \Delta\hat{p}\rangle+cst\end{equation}

\subsection{Phase locking and acquisition process}
\label{sub:lock}
Since our protocol relies on encoding information in the squeezing phase and on measuring precise elements of the covariance matrix, it is essential to stabilize both the pump phase and the local oscillator (LO) phase. The phase of the local oscillator $\theta_{LO}$ with respect to a fixed reference, that we conveniently call $\phi_S$, determines the quadrature measured via homodyne detection: $\hat{x}({\theta_{\text{LO}}}) = \cos(\theta_{LO})\hat{q} + \sin(\theta_{\text{LO}})\hat{p}$.

To achieve phase stabilization, a third beam, termed the seed, is introduced at the same wavelength as the LO. Since irrelevant during the actual squeezing measurements (which involve vacuum modes), the seed is periodically blocked during designated measurement times. In the alternating locking times, it is instead unblocked and used to actively monitor and stabilize the pump and LO phases. The detection is left inactive during the locking times, thanks to a gate function of the instruments (gating the measurement times). 

The seed follows the same optical path as the pump and interacts with it in the nonlinear waveguide via amplification and deamplification. These oscillations in the seed (and pump) amplitude are monitored and are proportional to:
\begin{equation}
  \cos(\phi_P - 2\phi_S)  
\end{equation}
where $\phi_P$ and $\phi_S$ are the pump and seed phases, respectively. This signal is used in a feedback loop via a Proportional – Integral – Derivative (PID) controller to lock the phase difference. During the locking times, the phase modulator used to encode the information is disabled via an analog switch, ensuring that the encoded phase difference $\delta$ remains unchanged (not compensated by the PID).

\begin{figure}
    \centering

        \includegraphics[width=\linewidth]{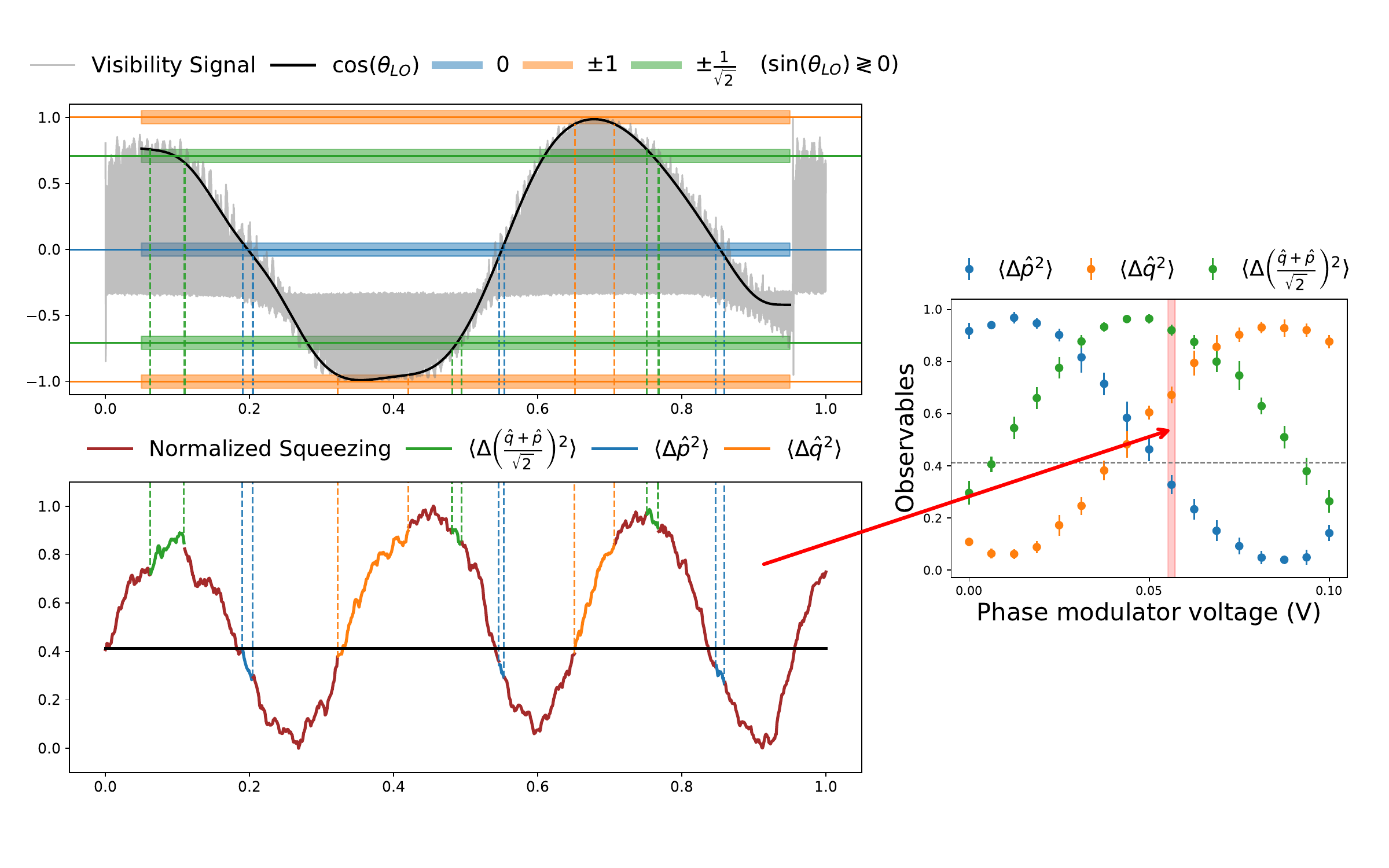}

    \caption{\textbf{Experimental acquisition process and phase dependence}. \textit{On the left} oscilloscope (top) and spectrum analyzer (bottom) normalized acquisitions for a single global pump phase. The visibility, acquired with the oscilloscope, is normalized between -1 and 1 as a cosine. The variance of the homodyne signal from the spectrum analyzer is normalized between 0 and 1 (to have already scaled observables). The values of the visibility cosine dictate what quadrature we are measuring ($\hat{x}_{\theta_{\text{LO}}} = \cos(\theta_{LO})\hat{q} + \sin(\theta_{\text{LO}})\hat{p}$), we take the corresponding times on the spectrum analyzer signal to reconstruct an average value of the covariance elements for the selected global phase (in dB). These averages become our observables. \textit{On the right} the results of these acquisitions when scanning the global phase of the pump with the modulator and by repeating $10$ times the acquisition on the left.}
    \label{fig:acquisition}
\end{figure}

The seed also interferes on its path with the LO, generating a signal known as visibility, which is sensitive to $\phi_{\text{LO}}$ because it is proportional to: \begin{equation}
    \cos(\phi_{\text{LO}}-\phi_S)=\cos{\theta_{LO}}
\end{equation}
 While this signal can be locked with a PID loop to a specific setpoint, corresponding to a given quadrature measurement, we opted for a more flexible approach. Instead of alternating multiple lock points, we scan the LO phase and post-select the squeezing measurements based on the value of the visibility signal, as illustrated in Fig.~\ref{fig:acquisition}. The visibility is seen on an oscilloscope from a leak after the balanced beam splitter of the homodyne detection, where seed and LO interact. The variance of the homodyne signal is measured with a spectrum analyzer, synchronized with the oscilloscope and the locking/measurement phases. Indeed, the spectrum analyzer calculates the variance of the input signal when used in zero-span, which gives us our observables directly. 

In our setup, the homodyne interferometer uses a local oscillator derived from the same laser that generates the quantum resource, which largely cancels common-mode noise. Balanced homodyne detection further suppresses residual fluctuations.

\subsection{Experimental noise}

Experimental instabilities introduce Gaussian noise in the measurement of observables. The main contributions are:

\begin{itemize}
    \item Mechanical instabilities: The setup uses an air-floated optical table with standard optomechanical mounts and 1-inch mirrors. The optical path is aligned at a global height of approximately 7.5 cm from the table surface. No specific vibration isolation has been implemented beyond precise mounting of the waveguide holders. A more compact and stable optomechanical setup could further reduce these fluctuations

  \item Residual phase instabilities: They are due to air diffusion (from air conditioning) in the room. The optical setup can be fully isolated by covering it.
    
    \item Detection and measurement protocol: Observables are recorded as detailed in Subsection\ref{sub:lock}. The calculation of variances along the scanned quadratures, as performed by the spectrum analyzer, is intrinsically noisy (see bottom panel in Fig. \ref{fig:acquisition}), as is the identification of quadratures with the visibility signal (Fig. \ref{fig:acquisition}). Improved measurement strategies, such as using the temporal trace recorded with an analog or digital oscilloscope to extract quadrature values pulse by pulse, including accurate electronic-noise filtering, together with more efficient and faster detectors, can further reduce the Gaussian noise.
\end{itemize}
Concerning the homodyne interferometer, the local oscillator is derived from the same laser that generates the quantum resource, which largely cancels common-mode noise.

\subsection{Parameters $\alpha, \beta$ and feedback mask}
\label{sub:par}
From the lower subplot of Fig.1 of the main text, we estimate that achieving the full modulation range of observables (from 0 to 1 or vice versa, so a $\delta$ span of $\pi/2$) requires a voltage span of approximately \( V_{\pi/2} = 0.075~\mathrm{V} \) (before amplification). This value sets the scale for the input and feedback parameters used across tasks. The global phase of the pump, $\delta$, depends indeed on the voltage $V$ applied to the EOM as $\delta(V) = \frac{\pi}{2} \cdot \frac{V}{V_{\pi/2}} + \text{offset}$.
\subsubsection*{XOR Task}

For the XOR task, the following parameters were used:
\begin{itemize}
    \item Input (\( s_k \in \{0,1\} \)) coefficient: \( \alpha = V_{\pi/2} \),
    \item Offset: \( \beta = -0.01 \) (a small offset with negligible effect on performance),
    \item Feedback coefficient (scalar): \( M = 0.035 \approx 0.46\, V_{\pi/2} \).
\end{itemize}

\subsubsection*{Memory Task}

No offset was applied for the memory task, i.e., \( \beta_r = 0 \) for all \( r \). The feedback strengths \( \alpha_r \), grouped in a vector \( \vec{\alpha} \in \mathbb{R}^R \), and the feedback mask matrices \( M_{r, r'} \in \mathbb{R}^{R \times R} \), were drawn randomly and scaled (multiplied) by \( V_{\pi/2} \). The specific values before scaling are:

\begin{itemize}
    \item For \( R = 5 \):
    \begin{align*}
        \vec{\alpha} &= \begin{bmatrix}
            -0.17 & -0.37 & -0.31 & -0.14 & -0.3
        \end{bmatrix}, \quad
        M = \begin{bmatrix}
            -0.39 & 0.16 & -0.34 & 0.17 & 0.32 \\
            0.45 & -0.21 & 0.13 & 0.43 & 0.32 \\
            -0.11 & 0.50 & 0.29 & 0.45 & 0.06 \\
            0.32 & 0.23 & -0.06 & -0.47 & 0.47 \\
            -0.26 & -0.01 & -0.17 & -0.12 & 0.33
        \end{bmatrix}
    \end{align*}
    
    \item For \( R = 3 \):
    \begin{align*}
        \vec{\alpha} &= \begin{bmatrix}
            0.25 & 0.34 & -0.32
        \end{bmatrix}, \quad
        M = \begin{bmatrix}
            0.33 & -0.02 & -0.34 \\
            -0.50 & -0.28 & 0.29 \\
            -0.40 & -0.48 & -0.35
        \end{bmatrix}
    \end{align*}
    
    \item For \( R = 1 \):
    \begin{align*}
        \vec{\alpha} &= \begin{bmatrix}
            -0.25
        \end{bmatrix}, \quad
        M = \begin{bmatrix}
            0.87
        \end{bmatrix}
    \end{align*}
\end{itemize}

\subsubsection*{Double Scroll Task}

For the double scroll task, no offset was applied. The feedback vector \( \vec{\alpha} \) was initialized randomly in the interval \( [-w, w] \) with \( w = 1.25 \). The feedback mask \( M \) was drawn from a uniform distribution over \( [-1, 1] \), normalized by its largest singular value, and scaled by a factor of 0.7. All values were then multiplied by \( V_{\pi/2} \).

\subsubsection*{General Encoding}

The general encoding used in all tasks included:
\begin{itemize}
    \item Delay time: \( \delta_{\text{amp}} = 10^{-16}~\mathrm{s} \),
    \item Random vectors \( \vec\alpha \) and \( \vec\beta \)  of length \( N \), with entries uniformly sampled from \( [-1, 1] \) and multiplied by \( \delta_{\text{amp}} \),
    \item Feedback mask matrix \( M \) with entries in \( [0, 2] \), normalized by its largest singular value, scaled by a factor \( m = 0.4 \) and multiplied by \( \delta_{\text{amp}} \).
\end{itemize}

\subsection{Double scroll ODEs}
\label{sub:ode}
\begin{figure}
    \centering
    \includegraphics[width=0.7\linewidth]{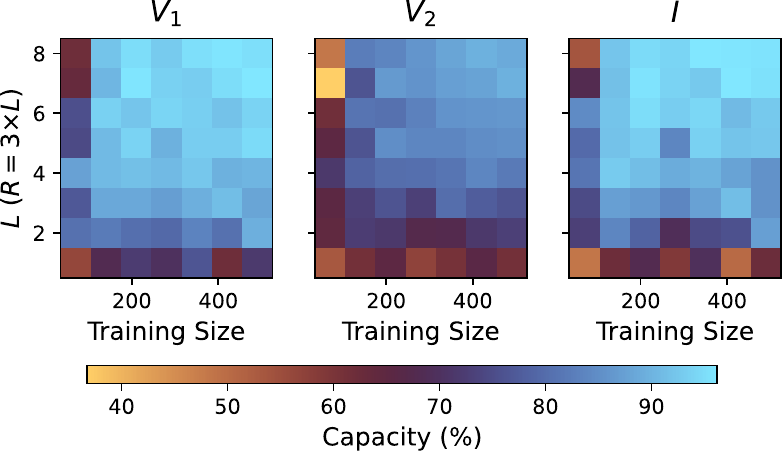}
    \caption{\textbf{Double-scroll task (simulated)}. Prediction capacity as a function of $L$ (number of reservoirs $=3L$) and training size, for $V_1$, $V_2$, and $I$. }
    \label{fig:double-scroll}
\end{figure}

 The system of ODEs to generate the data for the double-scroll task reads
\begin{equation}
\begin{aligned}
    \dot{V}_{1} &= V_{1}/R_{1} - \Delta V/R_{2} - 2 I_{r} \sinh(\beta \Delta V), \\
    \dot{V}_{2} &= \Delta V/R_{2} + 2I_{r} \sinh(\beta \Delta V) - I, \\
    \dot{I} &= V_{2} - R_{4}I \ ,
\end{aligned}
\end{equation}
where $\Delta V = V_{1} - V_{2}$, $(R_{1}, R_{2}, R_{4}) = (1.2, 3.44, 0.193)$, $\beta = 11.6$ and $I_{r} = 2.25 \times 10^{-5}$. We consider the input vector at each time step to be $\mathbf{s}_{k} = \left[ V_{1}(k\Delta t), V_{2}(k\Delta t), I(k\Delta t) \right]^{\top}$. In our case, the sampling interval is $\Delta t = 1$.

\subsection{Lorenz '63 ODEs}
\label{sub:lorenz}

To generate the input data for the Lorenz 63 task, we consider the standard Lorenz system of three coupled nonlinear differential equations:
\begin{equation}
\begin{aligned}
    \dot{x} &= \sigma (y - x), \\
    \dot{y} &= x (\rho - z) - y, \\
    \dot{z} &= x y - \beta z,
\end{aligned}
\end{equation}
where $\sigma$, $\rho$, and $\beta$ are the system parameters. In our simulations, we use the canonical chaotic regime values:
\begin{equation}
\sigma = 10.0, \quad \rho = 28.0, \quad \beta = \frac{8}{3}.
\end{equation}

The system is numerically integrated starting from the initial condition
\begin{equation}
(x_0, y_0, z_0) = (1.0, 1.0, 1.0),
\end{equation}
using a time increment $h$ (integration step) chosen as
\begin{equation}
h = 0.03 \, .
\end{equation}

\subsection{Comparison with other machine learning approaches}
\label{sec:comparison}

In this section, we compare our QRC performances with classical approaches and with another QRC reported in the literature \cite{wang2024enhanced}. 

While comparing quantum and classical approaches is crucial for identifying potential advantages, establishing a fair and meaningful benchmark between quantum and classical reservoirs or other machine learning approaches remains an open question. Differences in resources, encoding schemes, and dynamics make direct comparisons challenging. In practical terms, all the performances are strongly dependent on different resources (number of units, training and optimization time) as well as on the level of optimization of hyperparameters. In the following, we present results considering two well-known approaches in time series processing: Echo State Networks (ESN) and Long Short-Term Memory (LSTM). 

ESNs are a type of reservoir computers where the reservoir is a recurrent neural network (RNN) with random fixed weights. The tuning of the reservoir mainly relies on ensuring the echo state property (the input sequence should drive the dynamics of the reservoir regardless of its initial state) and on tuning memory capacity, through the spectral radius and other hyperparameters of the model. Only the readout is trained, consistently with the RC approach, with a linear regression. The number of neurons in the reservoir determines the dimensionality of the state vector and thus the number of degrees of freedom available to the linear readout. This makes ESNs a natural comparison point for our QRC, since both are reservoir computing approaches with fixed dynamics and trainable linear readout. In particular, we employ one of the simplest ESN models whose dynamics are given by:
\begin{equation}
    \mathbf{x}(t) = (1 - \alpha)\mathbf{x}(t-1) 
+ \alpha \tanh\left( 
\mathbf{W}_{\text{res}} \, \mathbf{x}(t-1) 
+ \mathbf{W}_{\text{in}} \, (\mathbf{u}(t) \odot s_{\text{scale}}) 
\right)
\label{eq:esn}
\end{equation}
where $\mathbf{x}(t) \in \mathbb{R}^{n}$ denotes the reservoir state at time $t$, 
$\mathbf{u}(t) \in \mathbb{R}^{d}$ is the input vector, 
$\mathbf{W_{\text{res}}} \in \mathbb{R}^{n' \times n'}$ is the recurrent (reservoir) weight matrix, $\mathbf{W}_{\text{in}} \in \mathbb{R}^{n' \times d}$ is the input weight matrix, and $\odot$ denotes element-wise multiplication. These connectivity matrices are randomly generated and remain fixed throughout the process. Here, $d$ and $n'$ represent the input dimension and the number of neurons in the reservoir.
The parameter $\alpha \in (0, 1]$ is the leaking rate that controls how fast the reservoir state updates, 
while $s_{\text{scale}}$ represents the input scaling factor. The final outcome of the ESN model is obtained by applying a linear layer to the reservoir state as features. The linear model used here is exactly the same as in the quantum case, trained by simple linear regression.  

LSTMs are a widely used neural network architecture for sequence learning. In contrast to ESNs or QRC, where the reservoir dynamics are fixed, an LSTM learns its internal dynamics during training. At each time step, it maintains a hidden state that is updated by trainable weights, and the output is computed from this hidden state. The internal structure of an LSTM comprises several hidden layers with different activation functions, along with a memory state. The LSTM uses a gate-based mechanism to control information flow: the forget gate decides what information to discard from the memory, the input gate determines which new information to store, and the output gate controls what information to pass to the next hidden state. More details on the LSTM structure can be found in \cite{LSTMpaper}. Furthermore, training optimizes both the update rules for the hidden state and the final mapping to the output, using gradient descent and backpropagation through time. The optimization is therefore significantly more computationally expensive than the other approaches, which use linear regression. The number of hidden units defines the dimension of the hidden state and thus the expressive capacity of the model, but it also determines the number of trainable parameters, which typically grows quadratically with this hidden size. LSTMs therefore represent a powerful state-of-the-art benchmark, though comparisons with QRC are not straightforward since their training complexity and parameter count are fundamentally different. In our case, a sequential TensorFlow model comprising one LSTM layer and a dense linear layer will be used \cite{tensorflow2015-whitepaper}. The architecture is built this way to create a structure similar to that of reservoir computing models, such that the LSTM layer serves as the reservoir and the final dense layer is the linear readout layer of the ESN and QRC models. Nonetheless, the full model is trained by employing the Adam optimization algorithm. 

\begin{figure}
    \centering
    \includegraphics[width=0.9\linewidth]{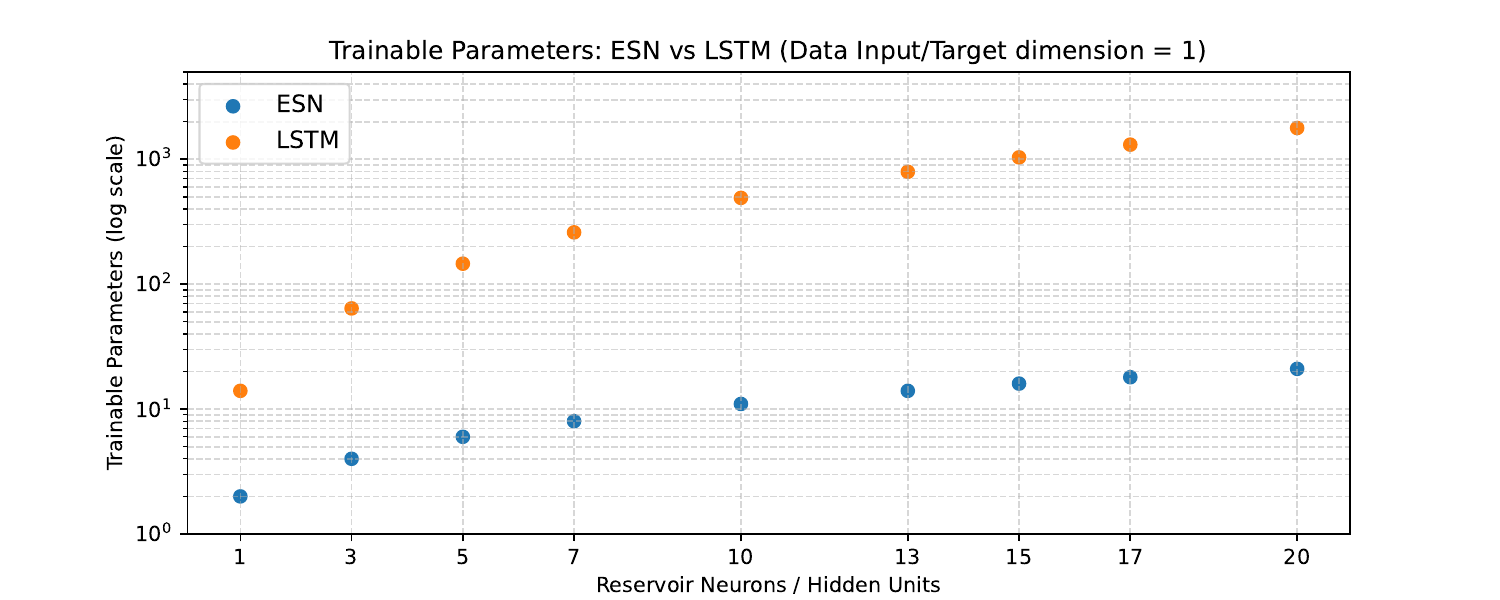}
    \caption{Comparison in the number of parameters to be trained for two classical approaches (Echo State Networks and Long Short-Term Memories) of time-series processing. The quantum model parameters would align with the one of the ESN. We can see that since the dynamics of LSTMs are themselves trained and not fixed like reservoirs, the amount of trainable parameters scales faster with the size. The plot shows trainable parameters for the input and output (target) dimensions equal to 1, corresponding to the XOR and memory task cases. In the forecasting tasks, the dimension would be 3, but the parameter scaling remains similar. The LSTM model parameters refer to the sum of the LSTM and dense layers employed in the architecture.}
    \label{fig:ressources}
\end{figure}

To compare equivalent QRC, ESN and LSTM settings one can constrain the dimension of the feature space used to approximate the target function or the number of neurons. For ESNs, the trainable parameters (feature space) correspond to the number of neurons; for LSTMs, to the number of hidden units and final output mapping; and for our QRC approach, to the number of observables extracted from the covariance matrix. However, from an experimental perspective, reconstructing $n(n+1)/2$ observables (for instance, the terms in $\hat{q}$ of a covariance matrix) requires measurements on only $n$ modes in their respective $\hat{q}$ quadratures. As shown in Fig. 4 of the main manuscript, by employing the general encoding we can obtain $n(n+1)/2$ independent observables while accessing only $n$ modes. This results in a polynomial scaling of available features with the number of modes, and we find it important to also present the comparison in terms of modes versus neurons/hidden units to highlight this polynomial advantage. Additionally, regarding LSTMs, since both the update rules for the hidden state and the final output mapping are trained, we found it more meaningful to compare LSTMs with ESNs having a similar number of trainable parameters rather than the same output layer size. The scaling of trainable parameters for ESNs and LSTMs is shown in Fig. \ref{fig:ressources}, highlighting the differences in resource requirements between these architectures.

We evaluated our QRC approach against classical reservoirs and LSTMs across multiple time-series tasks. For each task, the classical modes have been optimized via a grid search over their respective hyperparameters. In particular, for the ESN, we optimize for different values of $\alpha$, $s_{\text{scale}}$, sparsity, and spectral radius. For the LSTM, the number of epochs, learning rate, and dropout rate were adjusted. For LSTM training, a regularization term is not employed, since the linear regression we use for the reservoir computing models does not use one either. 

\begin{figure}
    \centering
    \includegraphics[width=\linewidth]{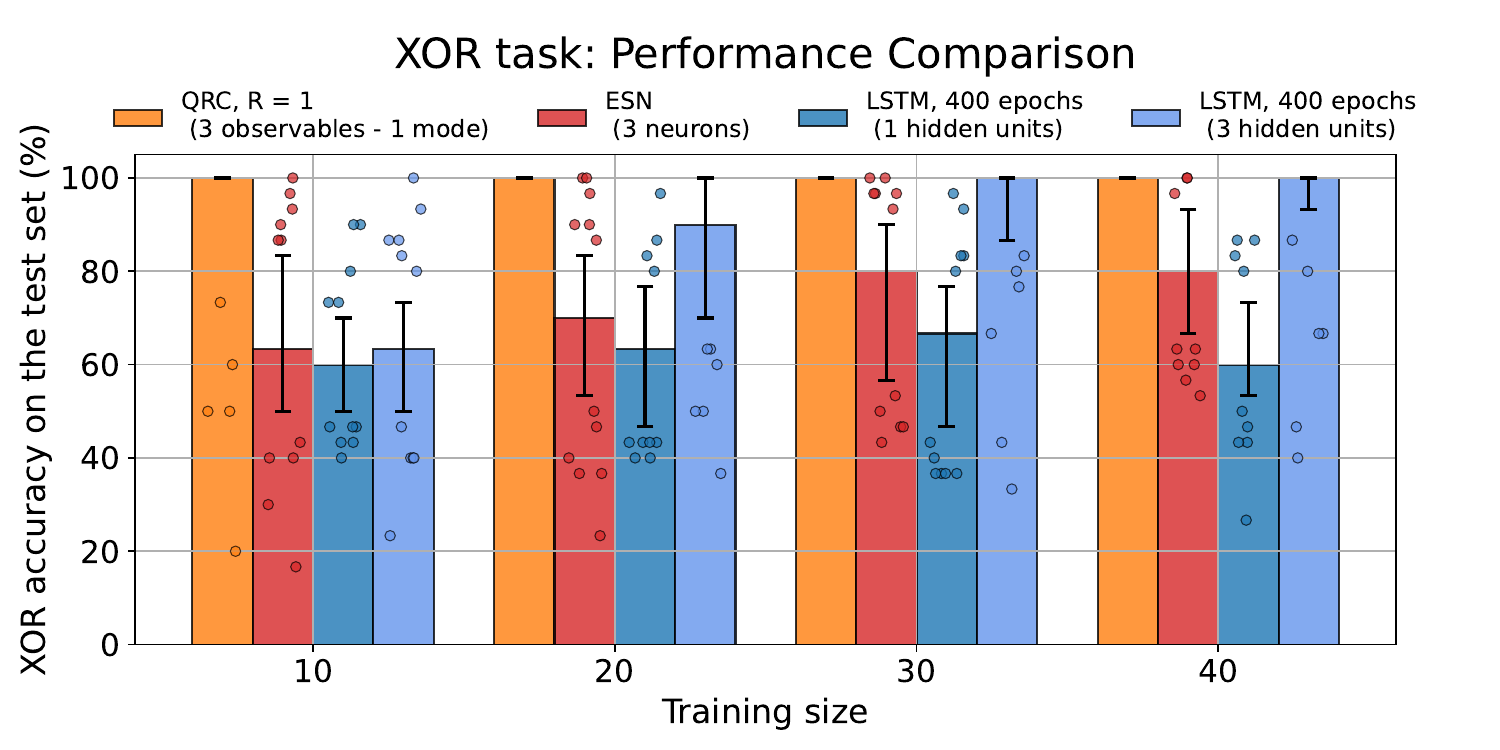}
    \caption{Comparison between our quantum reservoir computing approach (simulated noiseless global phase encoding, in orange) with the following classical time-series processing approaches: Echo State Networks (ESNs, in red) and Long Short-Term Memories (LSTMs, in two shades of blue according to the number of hidden units). The metric is the performance in a \textbf{temporal XOR} task, depending on the training size. The small training sizes affect more the classical approaches. The dimensions of the systems chosen have comparable output layer sizes or amount of trainable parameters. The bar height corresponds to the median of the trials. Error bars represent the 25th and 75th quartiles. Dots correspond to all data points outside this range. 
}
    \label{fig:XOR_vs_classical}
\end{figure}

\begin{figure}
    \centering
    \includegraphics[width=\linewidth]{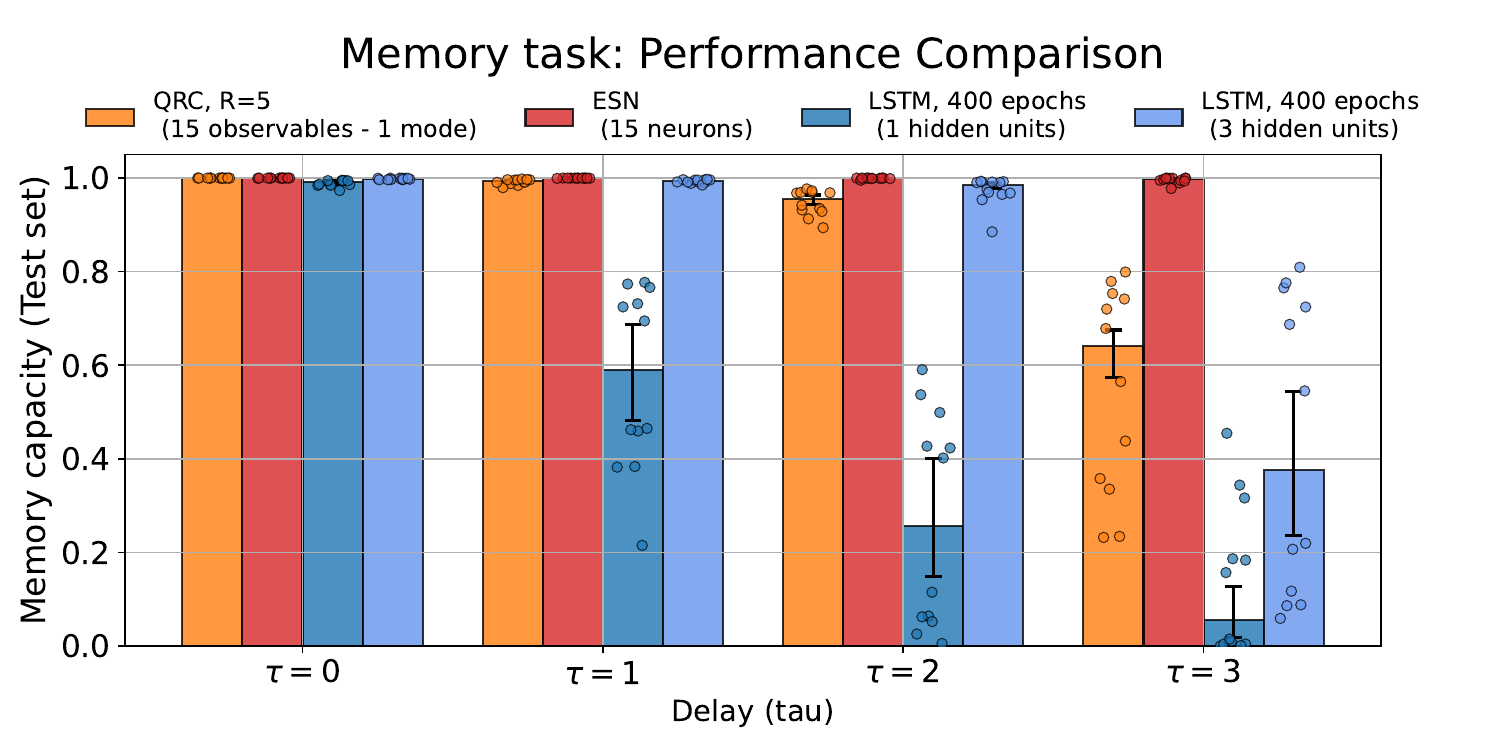}
    \caption{Comparison between our quantum reservoir computing approach (simulated noiseless global phase encoding, in orange) with the following classical time-series processing approaches: Echo State Networks (ESNs, in red) and Long Short-Term Memories (LSTMs, in two shades of blue according to the number of hidden units). The metric is the performance in a \textbf{Memory} task, depending on the delay $\tau$ to be remembered. Our system is less efficient than ESNs in this task, however compares well with LSTMs. The dimensions of the systems chosen have comparable output layer sizes or amount of trainable parameters. The bar height corresponds to the median of the trials. Error bars represent the 25th and 75th quartiles. Dots correspond to all data points outside this range.}
    \label{fig:memory_vs_classical}
\end{figure}

Figures~\ref{fig:XOR_vs_classical} and \ref{fig:memory_vs_classical} summarize the performance on the temporal XOR and linear memory tasks, respectively, already introduced in the manuscript.

In the temporal XOR task (Fig.~\ref{fig:XOR_vs_classical}), the QRC shows robust performance even for small training sizes, whereas classical approaches (ESNs and LSTMs) are more affected by limited training data. This aligns with results from previous work, which discuss the idea that quantum models can lead to better generalization with fewer training data \cite{caroGeneralizationQuantumMachine2022}. In the linear memory task (Fig.~\ref{fig:memory_vs_classical}), the QRC is less efficient than ESNs for short delays ($\tau$), but still compares favorably with LSTMs when the amount of trainable parameters is similar. It was expected that the ESN could achieve better performance in this linear task, since the trade-off between linear and nonlinear dynamics is regulated by its hyperparameter $\alpha$. In the quantum case, tuning between linear and nonlinear is not so straightforward, leading to lower performance compared to the ESN for this task.

\begin{figure}
    \centering

        \includegraphics[width=\linewidth]{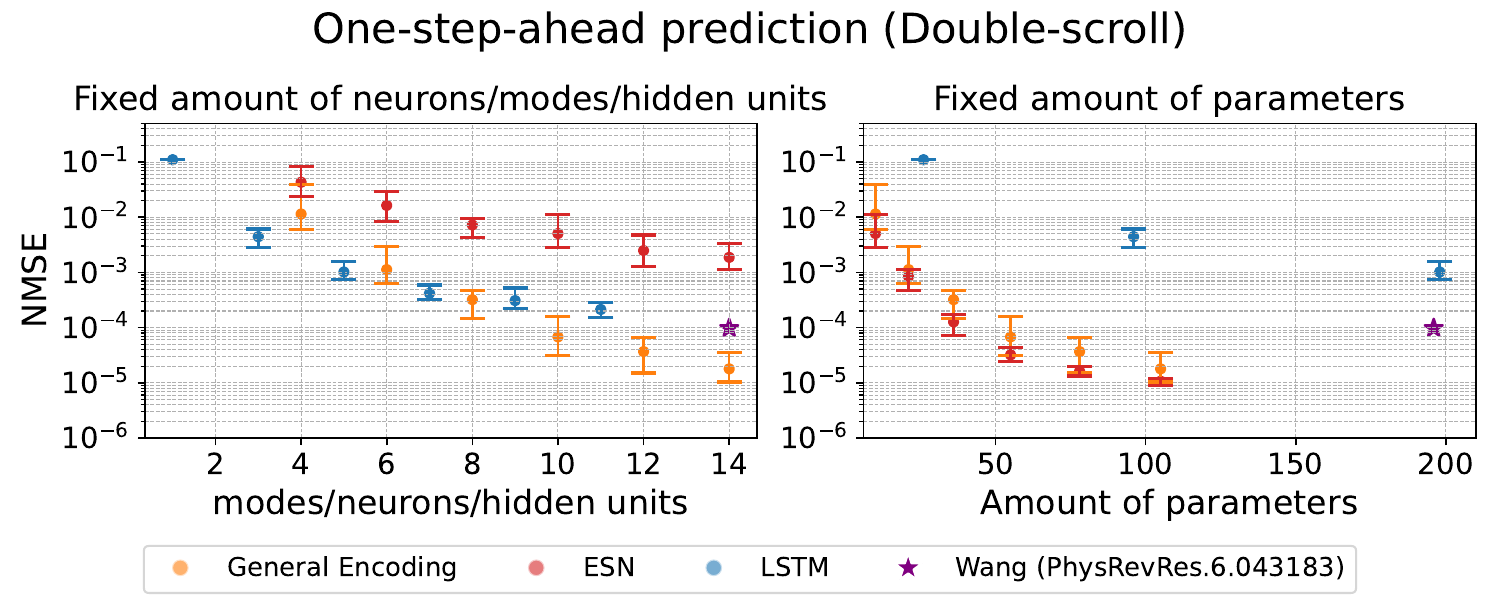}
    \\[0.5cm]

        \includegraphics[width=\linewidth]{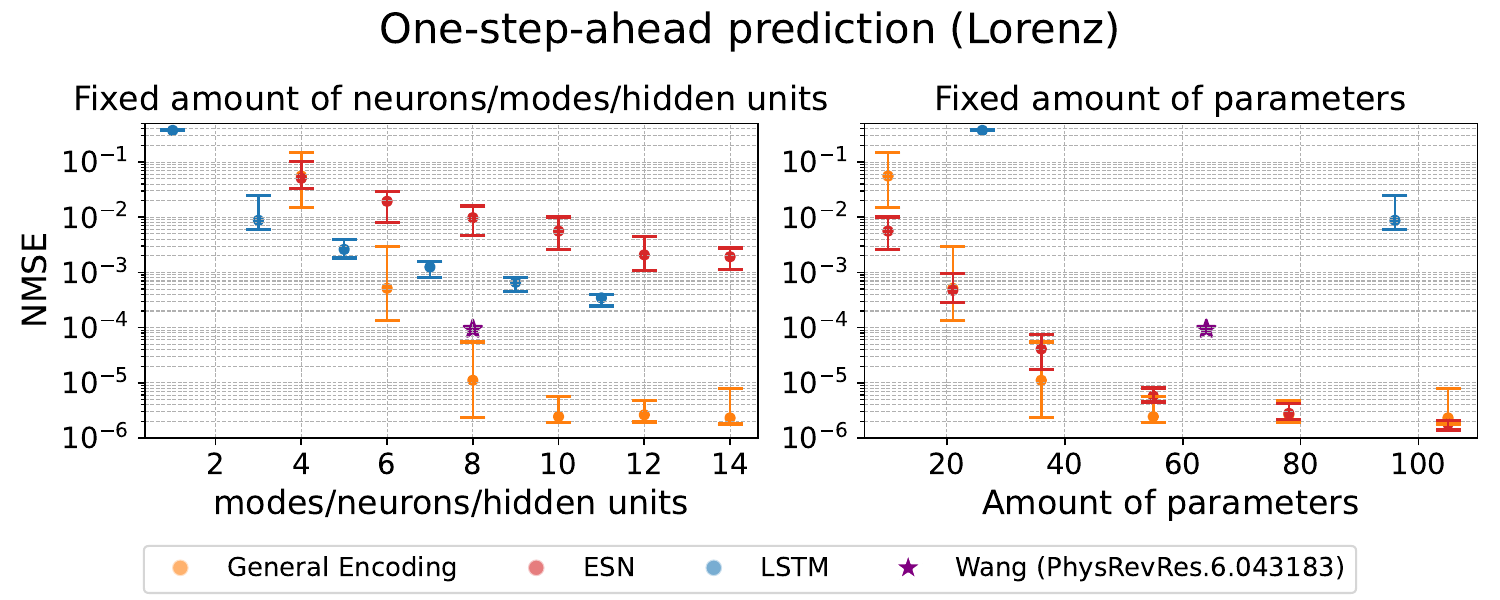}
    
    \caption{Comparison between our quantum reservoir computing approach (simulated noiseless general encoding, in orange) with the following classical time-series processing approaches: Echo State Networks (ESNs, in red), Long Short-Term Memories (LSTMs, in blue), and another quantum reservoir computing approach} (\cite{wang2024enhanced}, in purple). The metric is the performance in one-step-ahead prediction for two different temporal series tasks: \textbf{Double-scroll} on the top and \textbf{Lorenz} on the bottom. We present two different comparisons: on the left, equalizing the number of modes measured with the number of neurons/hidden units, which allows us to showcase for the quantum reservoir the quadratic scaling of the number of observables with the number of modes $n$ (visible when comparing with ESNs); and on the right equalizing the number of parameters to be trained, where the quantum reservoir achieves performance comparable to the ESNs and superior to the LSTMs. The y-values of the dots correspond to the median of the trials, and error bars represent the 15th and 85th percentiles. For both tasks, the training size was 2500 and the test size 250.
    \label{fig:double_lorenz_vs_classical}
\end{figure}

Figure~\ref{fig:double_lorenz_vs_classical} shows one-step-ahead prediction performance on the double-scroll and Lorenz chaotic systems, respectively. We report two types of comparisons for each task: one matching the number of modes to neurons/hidden units and one matching the number of trainable parameters. The first one highlights the polynomial scaling of available observables in QRC. Measuring $n$ modes allows access to $n(n+1)/2$ independent observables, which outperforms classical reservoirs of similar size (visible when comparing with ESNs). In the second one, QRC is comparable to ESNs in terms of parameter count but has an advantage over LSTMs due to the latter’s more complex training of hidden dynamics. Note that in these plots, the points on the left and right correspond to the same data for the LSTM and quantum model, but the x-axis is scaled by the number of parameters. However, the data for the ESN in the left and right plots differ, as it was necessary to increase the number of neurons to match the trainable parameters used with the quantum model. 

We also include a comparison to another recent quantum reservoir computing approach for reference \cite{wang2024enhanced}.

These results clearly illustrate the efficiency of QRC in leveraging fixed quantum dynamics to generate a high-dimensional feature space with polynomial scaling, while maintaining competitive predictive performance with classical benchmarks.

\subsection{Summary of performances of our QRC approach}

Table~\ref{tab:performance_summary} summarizes the tasks considered in this work, including their inputs, target outputs, and the corresponding performance metrics used to evaluate the quantum reservoir.

\begin{sidewaystable*}[t]
\centering
\renewcommand{\arraystretch}{1.3}
\setlength{\tabcolsep}{4pt} 
\small 

\begin{tabular}{|l|l|l|l|l|}
\hline
\textbf{Task} & \textbf{Input} & \textbf{Target output $\hat{y}_k$} & \textbf{Performance metric} & \textbf{Performance on the test set} \\ \hline

\textit{Temporal XOR} & Binary sequence $s_k \in \{0,1\}$ &
$\hat{y}_k = s_k \oplus s_{k-1}$ &
\parbox[t]{4.8cm}{After thresholding predictions at 0.5:\\ $\mathit{Accuracy}= \frac{1}{M} \sum_{i=1}^{M} \delta(y'_i, \hat{y}_i)$} &
\parbox[t]{5.7cm}{$98\pm1$\%\\ (\textbf{global phase encoding - experimental realization with the smallest noise}, 70 training steps)} \\ \hline

\textit{Linear memory} & Continuous seq. $s_k \in [-1,1]$ &
$\hat{y}_k = s_{k-\tau}$ &
\parbox[t]{4.8cm}{$\mathit{Capacity} = (\mathrm{corr}(y, \hat{y}))^2 = \left( \frac{\mathrm{Cov}(y, \hat{y})}{\sigma_y \sigma_{\hat{y}}} \right)^2$} &
\parbox[t]{5.7cm}{$0.93\pm0.03$ ($\tau=0$)\\ $0.58\pm0.1$ ($\tau=1$)\\ $0.14\pm0.1$ ($\tau=2$)\\ $0.04\pm0.06$ ($\tau=3$)\\ (\textbf{global phase encoding - experimental realization with R=5 multiplexing}, 99 training steps)} \\ \hline

\textit{Double scroll} & 3D continuous seq. $\mathbf{s}_k$ (\ref{sub:ode}) &
$\hat{\mathbf{y}}_k = \mathbf{s}_{k+1}$ &
\parbox[t]{4.8cm}{$\mathit{Capacity} = (\mathrm{corr}(\mathbf{y}, \hat{\mathbf{y}}))^2 = \left( \frac{\mathrm{Cov}(\mathbf{y}, \hat{\mathbf{y}})}{\sigma_{\mathbf{y}} \sigma_{\hat{\mathbf{y}}}} \right)^2$} &
\parbox[t]{5.7cm}{$96.1\pm0.6$\% ($V_1$)\\ $90.0\pm2.3$\% ($V_2$)\\ $95.4\pm0.9$\% ($I$)\\ (\textbf{global phase encoding - numerical simulation with experimental noise}, R=24 multiplexing, 350 training steps)} \\ \hline

\textit{Parity check} & Binary seq. $s_k \in \{0,1\}$ &
$\hat{y}_k = \sum_{i=k-\tau}^{k} s_i \bmod 2$ &
\parbox[t]{4.8cm}{$\mathit{Accuracy} = \frac{1}{M} \sum_{i=1}^{M} \delta(y'_i, \hat{y}_i)$} &
\parbox[t]{5.7cm}{$100\pm0$\% ($\tau=0$–$3$)\\ $73\pm16$\% ($\tau=4$)\\ $52\pm8$\% ($\tau=5$)\\ (\textbf{general encoding - numerical simulation}, N=5, n=6, 180 training steps)} \\ \hline

\textit{Double scroll} & 3D continuous seq. $\mathbf{s}_k$ (\ref{sub:ode}) &
$\hat{\mathbf{y}}_k = \mathbf{s}_{k+1}$ &
\parbox[t]{4.8cm}{$\mathit{NMSE} = \frac{\langle (y - \hat{y})^2 \rangle}{\langle y^2 \rangle}$ (all 3 dimensions)} &
\parbox[t]{5.7cm}{NMSE: $1.24\times10^{-6}$\\ NMSE std: $4.74\times10^{-6}$\\ (\textbf{general encoding - numerical simulation}, $N=15$, $n=14$, 2500 training steps)} \\ \hline

\textit{Lorenz} & 3D continuous seq. $\mathbf{s}_k$ (\ref{sub:lorenz}) &
$\hat{\mathbf{y}}_k = \mathbf{s}_{k+1}$ &
\parbox[t]{4.8cm}{$\mathit{NMSE} = \frac{\langle (y - \hat{y})^2 \rangle}{\langle y^2 \rangle}$ (all 3 dimensions)} &
\parbox[t]{5.7cm}{NMSE: $7.01\times10^{-6}$\\ NMSE std: $3.97\times10^{-5}$\\ (\textbf{general encoding - numerical simulation}, $N=15$, $n=14$, 2500 training steps)} \\ \hline

\end{tabular}
\caption{Summary of best performances across tasks.}
\label{tab:performance_summary}
\end{sidewaystable*}

\end{document}